%%%%%%%%%%%%%%%%%%%%%%%%%%%%%%%%%%%%%%%%%%%%%%%%%%%%%%%%%%%%%%

\input harvmac
%\draftmode
\def \lc {light-cone\ }

\def \const {{\rm const}}
\def \la {\langle}

\def \ov {\over}

\def \ep {\epsilon}
\def \k {\kappa}
\def \N {{\cal N}}
\def  \cN {{\cal N}}
\def  \rN {{\rm N}}
\def \cH {{\cal H}}

\def \K {{\cal K}}

\def \S {{\cal S}}
\def \ss {{\cal S}}

\def \a {\alpha}
\def \E {{\cal E}}
\def \b {\beta}
\def \g {\gamma}
\def \G {\Gamma}

\def \l {\lambda}

\def \m {\mu}
\def \n {\nu}

\def \s {\sigma}
\def \ee {\epsilon}

\def \r {\rho}
\def \t {\theta}

\def \p {\phi}
\def \P { \Phi}

\def \vp {\varphi}

\def \frac#1#2{{ #1 \over #2}}
\def \lr { \lref}
\def \td {\tilde}

\def \lr{\lref}

\def \rf {\refs}

\def \ee {{\rm e}}

\def \adss {$AdS_5 \times S^5\ $}

\def \N {{\cal N}}
\def \lc {light-cone\ }

\def \s { \sigma }

\def \vp {\varphi}

\def \p {\phi}
\def \vt {\theta} \def \vtt{\vartheta}
 \def \a { \alpha}
\def \r {\rho}
\def \fourth {{1 \ov 4}}

\def \DD {{\cal D}}

\def \DD {{\rm D}}
\def \vr {\varrho}
  \def \td { \tilde }

\def \t {\theta}
\def \la {\label}

\def \del{\partial}
\def \m {\mu }
\def \n {\nu }
\def \ha { { 1 \over 2}}
\def \cN {{\cal N}}

\def \g {\gamma}
\def \G {\Gamma}
\def \k {\kappa}
\def \l {\lambda}

\def \td {\tilde }
\def \b{\beta}

\def \la {\label}

\def \ha {{1 \over 2}}
\def \ep{\epsilon}

\def \ov {\over}
 \def \Om {\Omega}

\def \om {\omega}

\def \sql {{\sqrt{\l}}}

\def \rr {{\td \r}}
\def \D {{\rm D}}
\def \E {{\cal E}}

\def \E  {{\cal E}}
\def \D {{\rm D}}\def \K {{\rm K}}
\def \P {{\rm P}}

\def \PP {{\cal P}}
\def \DD {{\cal D}} \def \KK {{\cal K}}

\def \t {\tau} 
\def \rf {\refs}
\def \T  {{\rm T}}

\def \thet {\theta} 

\def \om {\nu}
\def \ee {\epsilon}

\def \ss { s}

\def \rN {{\rm N}}

\def \ton {T^{1,1}} 

%%%%%%%%%%%%%%%%%%%%%%%%%%%%%%%%%%%%%%%%%%%%%%%%%%%%%%%%%%%%%%
\lr \zan {
A.~Santambrogio and D.~Zanon,
'`Exact anomalous dimensions of N = 4 Yang-Mills operators with large R  charge,''
hep-th/0206079.
%%CITATION = HEP-TH 0206079;%% 
}

\lr \mtt {
R.~R.~Metsaev and A.~A.~Tseytlin,
``Superstring action in \adss: kappa-symmetry light cone gauge,''
Phys.\ Rev.\ D {\bf 63}, 046002 (2001)
[hep-th/0007036].
%%CITATION = HEP-TH 0007036;%%
R.~R.~Metsaev, C.~B.~Thorn and A.~A.~Tseytlin,
``Light-cone superstring in AdS space-time,''
Nucl.\ Phys.\ B {\bf 596}, 151 (2001)
[hep-th/0009171].
%%CITATION = HEP-TH 0009171;%%
}

\lr \frt {
S.~Frolov and A.~A.~Tseytlin,
``Semiclassical quantization of rotating superstring in \adss,''
JHEP {\bf 0206}, 007 (2002)
[hep-th/0204226].
%%CITATION = HEP-TH 0204226;%%
} 

\lr \mal {J.~M.~Maldacena,
``The large $N$ limit of superconformal field theories and supergravity,''
Adv.\ Theor.\ Math.\ Phys.\  {\bf 2}, 231 (1998)
[Int.\ J.\ Theor.\ Phys.\  {\bf 38}, 1113 (1999)]
[hep-th/9711200].
%%CITATION = HEP-TH 9711200;%%
O.~Aharony, S.~S.~Gubser, J.~M.~Maldacena, H.~Ooguri and Y.~Oz,
``Large N field theories, string theory and gravity,''
Phys.\ Rept.\  {\bf 323}, 183 (2000)
[hep-th/9905111].
%%CITATION = HEP-TH 9905111;%%
}

\lr \nast { D.~Berenstein and H.~Nastase,
``On lightcone string field theory from super Yang-Mills and holography,''
hep-th/0205048.
%%CITATION = HEP-TH 0205048;%%
}

\lr \parn{
A.~Parnachev and A.~V.~Ryzhov,
``Strings in the near plane wave background and AdS/CFT,''
hep-th/0208010.
%%CITATION = HEP-TH 0208010;%%
}

\lr \tsett{
A. Tseytlin, talk at KIAS Workshop on Strings and Branes, Seoul,  May 20-31, 2002 
http://m.kias.re.kr/program/program.htm;
talk at Strings 2002, Cambridge, July 15-20, 2002, 
http://www.damtp.cam.ac.uk/strings02/avt/tseytlin/
 }

\lr \vega { H.~J.~de Vega and I.~L.~Egusquiza,
``Planetoid String Solutions in 3 + 1 Axisymmetric Spacetimes,''
Phys.\ Rev.\ D {\bf 54}, 7513 (1996)
[hep-th/9607056].
%%CITATION = HEP-TH 9607056;%%
}

 \lr \ves { H.~J.~de Vega, A.~L.~Larsen and N.~Sanchez,
``Semiclassical quantization of circular strings in de Sitter and
anti-de Sitter space-times,''
Phys.\ Rev.\ D {\bf 51}, 6917 (1995)
[hep-th/9410219].
%%CITATION = HEP-TH 9410219;%%
}

\lr \tss   { A.~A.~Tseytlin,
``On limits of superstring in \adss,''
hep-th/0201112.
%%CITATION = HEP-TH 0201112;%%
}

\lr \poly {A. Polyakov, unpublished  (private communication, Jan. 2000). }

\lr \bmn { D.~Berenstein, J.~Maldacena and H.~Nastase,
``Strings in flat space and pp waves from N = 4 super Yang Mills,''
JHEP {\bf 0204}, 013 (2002)
[hep-th/0202021].
%%CITATION = HEP-TH 0202021;%%
}

\lr \gkp { S.~S.~Gubser, I.~R.~Klebanov and A.~M.~Polyakov,
``A semi-classical limit of the gauge/string correspondence,''
Nucl.\ Phys.\ B {\bf 636}, 99 (2002) [hep-th/0204051].
%%CITATION = HEP-TH 0204051;%%
}

\lr \blau{M.~Blau, J.~Figueroa-O'Farrill, C.~Hull and G.~Papadopoulos,
``A new maximally supersymmetric background of IIB superstring theory,''
JHEP {\bf 0201}, 047 (2002)
[hep-th/0110242].
%%CITATION = HEP-TH 0110242;%%
``Penrose limits and maximal supersymmetry,''
hep-th/0201081.
%%CITATION = HEP-TH 0201081;%%
}
\lr \malrey {J.~Maldacena,
``Wilson loops in large N field theories,''
Phys.\ Rev.\ Lett.\  {\bf 80}, 4859 (1998)
[hep-th/9803002].
%%CITATION = HEP-TH 9803002;%%
S.~J.~Rey and J.~Yee,
``Macroscopic strings as heavy quarks in large N gauge theory and
anti-de Sitter supergravity,''
Eur.\ Phys.\ J.\ C {\bf 22}, 379 (2001)
[hep-th/9803001].
%%CITATION = HEP-TH 9803001;%%
}

\lref\mets{
R.~R.~Metsaev and A.~A.~Tseytlin,
``Exactly solvable model of superstring in plane wave Ramond-Ramond
background,''
Phys.\ Rev.\ D {\bf 65}, 126004 (2002)
[hep-th/0202109].
%%CITATION = HEP-TH 0202109;%%
}

\lr \poll{A. Polyakov, talk at Strings 2002, 
%Cambridge, July 15-20, 2002, 
www.damtp.cam.ac.uk/strings02/avt/polyakov/ }

\lr \KS {
I.~R.~Klebanov and M.~J.~Strassler,
``Supergravity and a confining gauge theory: Duality cascades and  chiSB-resolution of naked singularities,''
JHEP {\bf 0008}, 052 (2000)
[hep-th/0007191].
%%CITATION = HEP-TH 0007191;%%
}

\lr \polk {A.~M.~Polyakov,
``Gauge fields and space-time,''
hep-th/0110196.
%%CITATION = HEP-TH 0110196;%%
}

\lref\met{
R.~R.~Metsaev,
``Type IIB Green-Schwarz superstring in plane wave Ramond-Ramond
background,''
Nucl.\ Phys.\ B {\bf 625}, 70 (2002)
[hep-th/0112044].
%%CITATION = HEP-TH 0112044;%%
}

\lref\MET{
R.~R.~Metsaev and A.~A.~Tseytlin,
``Type IIB superstring action in \adss  background,''
Nucl.\ Phys.\ B {\bf 533}, 109 (1998)
[hep-th/9805028].
%%CITATION = HEP-TH 9805028;%%
}
\lr \dgt {N.~Drukker, D.~J.~Gross and A.~A.~Tseytlin,
``Green-Schwarz string in \adss : Semiclassical partition
function,''
JHEP {\bf 0004}, 021 (2000)
[hep-th/0001204].
%%CITATION = HEP-TH 0001204;%%
A.~A.~Tseytlin,
``'Long' quantum superstrings in \adss ,''
hep-th/0008107.
%%CITATION = HEP-TH 0008107;%%
}
\lref\kat{
R.~Kallosh and A.~A.~Tseytlin,
``Simplifying superstring action on \adss,''
JHEP {\bf 9810}, 016 (1998)
[hep-th/9808088].
%%CITATION = HEP-TH 9808088;%%
}

\lr \thei  { S.~Forste, D.~Ghoshal and S.~Theisen,
``Stringy corrections to the Wilson loop in N = 4 super Yang-Mills
theory,''
JHEP {\bf 9908}, 013 (1999)
[hep-th/9903042].
%%CITATION = HEP-TH 9903042;%%
}

\lr \pol {A.~M.~Polyakov,
``Gauge fields and space-time,''
hep-th/0110196.
%%CITATION = HEP-TH 0110196;%%
}

\lr \sfets {  K.~Sfetsos and A.~A.~Tseytlin,
``Four-dimensional plane wave string solutions with coset CFT description,''
Nucl.\ Phys.\ B {\bf 427}, 245 (1994)
[hep-th/9404063].
%%CITATION = HEP-TH 9404063;%%
}

\lr \others {A.~Armoni, J.~L.~Barbon and A.~C.~Petkou,
``Orbiting strings in AdS black holes and N = 4 SYM at finite  temperature,''
JHEP {\bf 0206}, 058 (2002)
[hep-th/0205280].
%%CITATION = HEP-TH 0205280;%%
G.~Mandal, N.~V.~Suryanarayana and S.~R.~Wadia,
``Aspects of semiclassical strings in AdS(5),''
Phys.\ Lett.\ B {\bf 543}, 81 (2002)
[hep-th/0206103].
%%CITATION = HEP-TH 0206103;%%
}

\lref\min{
J.~A.~Minahan,
``Circular Semiclassical String Solutions on \adss,''
hep-th/0209047.
%%CITATION = HEP-TH 0209047;%%
}

\lref\rus{
J.~G.~Russo,
``Anomalous dimensions in gauge theories from rotating strings in \adss,''
JHEP {\bf 0206}, 038 (2002)
[hep-th/0205244].
%%CITATION = HEP-TH 0205244;%%
}

\lref\gro{
D.~J.~Gross, A.~Mikhailov and R.~Roiban,
``Operators with large R charge in N = 4 Yang-Mills theory,''
hep-th/0205066.
%%CITATION = HEP-TH 0205066;%%
}

\lr \old{
D.~J.~Gross and F.~Wilczek,
``Asymptotically Free Gauge Theories. 2,''
Phys.\ Rev.\ D {\bf 9}, 980 (1974).
%%CITATION = PHRVA,D9,980;%%
A.~Gonzalez-Arroyo and C.~Lopez,
``Second Order Contributions To The Structure Functions In Deep Inelastic Scattering. 3. The Singlet Case,''
Nucl.\ Phys.\ B {\bf 166}, 429 (1980).
%%CITATION = NUPHA,B166,429;%%
G.~P.~Korchemsky and G.~Marchesini,
``Structure function for large x and renormalization of Wilson loop,''
Nucl.\ Phys.\ B {\bf 406}, 225 (1993)
[hep-ph/9210281].
%%CITATION = HEP-PH 9210281;%%
}

\lr \ruslan{
R.~R.~Metsaev,
``Light cone gauge formulation of IIB supergravity in AdS(5) x S(5)  background and AdS/CFT correspondence,''
Phys.\ Lett.\ B {\bf 468}, 65 (1999)
[hep-th/9908114].
%%CITATION = HEP-TH 9908114;%%
}

\lr \plef {C.~Kristjansen, J.~Plefka, G.~W.~Semenoff and M.~Staudacher,
``A new double-scaling limit of N = 4 super Yang-Mills theory and PP-wave  strings,''
hep-th/0205033.
%%CITATION = HEP-TH 0205033;%%
N.~R.~Constable, D.~Z.~Freedman, M.~Headrick, S.~Minwalla, L.~Motl, A.~Postnikov and W.~Skiba,
``PP-wave string interactions from perturbative Yang-Mills theory,''
JHEP {\bf 0207}, 017 (2002)
[hep-th/0205089].
%%CITATION = HEP-TH 0205089;%%
}

\lr \rru{ J. Russo, unpublished.} 

\lr \dash{  R.~F.~Dashen, B.~Hasslacher and A.~Neveu,
``The Particle Spectrum In Model Field Theories From Semiclassical Functional Integral Techniques,''
Phys.\ Rev.\ D {\bf 11}, 3424 (1975).
%%CITATION = PHRVA,D11,3424;%%
}

%%%%%%%%%%%%%%%%%%%%%%%%%%%%%%%%%%%%%%%%%%%%%%%%%%%%%%%%%%%%%%%%%%%%%
\lr \KT { I.~R.~Klebanov and A.~A.~Tseytlin,
``Gravity duals of supersymmetric SU(N) x SU(N+M) gauge theories,''
Nucl.\ Phys.\ B {\bf 578}, 123 (2000)
[hep-th/0002159].
%%CITATION = HEP-TH 0002159;%%
}

\lr \kll {N.~Itzhaki, I.~R.~Klebanov and S.~Mukhi,
``PP wave limit and enhanced supersymmetry in gauge theories,''
JHEP {\bf 0203}, 048 (2002)
[hep-th/0202153].
%%CITATION = HEP-TH 0202153;%%
}
\lr \oog {
J.~Gomis and H.~Ooguri,
'`Penrose limit of N = 1 gauge theories,''
hep-th/0202157.
%%CITATION = HEP-TH 0202157;%%
}
\lr \ppz { 
L.~A.~Pando Zayas and J.~Sonnenschein,
'`On Penrose limits and gauge theories,''
JHEP {\bf 0205}, 010 (2002)
[hep-th/0202186].
%%CITATION = HEP-TH 0202186;%%
}

\lr \gimo { 
E.~G.~Gimon, L.~A.~Pando Zayas and J.~Sonnenschein,
'`Penrose limits and RG flows,''
hep-th/0206033.
%%CITATION = HEP-TH 0206033;%%
}

\lr \lei {
R.~Corrado, N.~Halmagyi, K.~D.~Kennaway and N.~P.~Warner,
'`Penrose limits of RG fixed points and pp-waves with background fluxes,''
hep-th/0205314.
%%CITATION = HEP-TH 0205314;%%
}
\lr \bre {
D.~Brecher, C.~V.~Johnson, K.~J.~Lovis and R.~C.~Myers,
``Penrose limits, deformed pp-waves and the string duals of N = 1 large N  gauge theory,''
hep-th/0206045.
%%CITATION = HEP-TH 0206045;%%
}

\lr \fkt {S.~Frolov, I.~R.~Klebanov and A.~A.~Tseytlin,
``String corrections to the holographic RG flow of supersymmetric  SU(N) x SU(N+M) gauge theory,''
Nucl.\ Phys.\ B {\bf 620}, 84 (2002)
[hep-th/0108106].
%%CITATION = HEP-TH 0108106;%%
}

\lr \KW{
I.~R.~Klebanov and E.~Witten,
``Superconformal field theory on threebranes at a Calabi-Yau  singularity,''
Nucl.\ Phys.\ B {\bf 536}, 199 (1998)
[hep-th/9807080].
%%CITATION = HEP-TH 9807080;%%
}

\lr \LS {  R.~G.~Leigh and M.~J.~Strassler,
``Exactly marginal operators and duality in four-dimensional N=1 supersymmetric gauge theory,''
Nucl.\ Phys.\ B {\bf 447}, 95 (1995)
[hep-th/9503121].
%%CITATION = HEP-TH 9503121;%%
}

%%%%%%%%%%%%%%%%%%%%%%%%%%%%%%%%%%%%%%%%%%%%%%%%%%%%%
\Title{\vbox
{\baselineskip 10pt
{\hbox{  Imperial/TP/1-02 /28 }
}}}
{\vbox{\vskip -30 true pt
\centerline {Semiclassical  quantization of
 superstrings: }
\medskip
\centerline {   $AdS_5 \times S^5$  and beyond }
\medskip
\vskip4pt }}
\vskip -20 true pt
\centerline{ A.A. Tseytlin$^{a,b,}$\footnote{$^{*}$}
{Also at
Lebedev Physics Institute, Moscow.}
}
\smallskip\smallskip
\centerline{ $^a$ \it  Blackett Laboratory,
 Imperial College,
 London,  SW7 2BZ, U.K.}

\centerline{ $^b$ \it  Department of Physics,
 The Ohio State University,
 Columbus, OH 43210, USA}

\bigskip\bigskip
\centerline {\bf Abstract}
\baselineskip12pt
\noindent
\medskip
We discuss  semiclassical quantization of closed  superstrings  in $AdS_5 \times S^5$. 
%following hep-th/0204226.
We consider  two  basic examples: point-like string boosted along large 
circle of  $S^5$  and  folded  string rotating in  $AdS_5$.
In the first  case  we clarify the general 
structure  of the sigma model  perturbation theory  for the energy of string states 
beyond the 1-loop order (related to the plane-wave limit). 
In the  second  case we argue that  the large  spin 
limit of the  expression for the ground-state 
energy (i.e.  for the  dimension of the   corresponding
minimal twist  gauge theory operator) 
has  the form $S + f(\l) \ln S$ 
to all orders in the $\a' \sim {1 \ov \sql}$ expansion, 
in agreement with  the AdS/CFT  duality. 
We also suggest  the  extension  of the semiclassical  approach to 
near-conformal (near-AdS)  cases  on the example  of the 
fractional D3-brane on conifold  background.

%{\centerline {\it Expanded version of  talks at  the Third International 
%Sakharov Conference on Physics,}}
%{\centerline {\it  Moscow,  June  24-29, 2002
%and at Strings 2002, Cambridge,   July 15-20,2002} }

%\bigskip
%\bigskip
\bigskip
\bigskip

%%%%%%%%%%%%%%%%%%%%%%%%%%%%%%%%%%%%%%%%%%%%%%%%%%%%%%%%%

%{\centerline {\it Expanded version of  talks at  the Third International 
%Sakharov Conference on Physics,}}
%{\centerline {\it  Moscow,  June  24-29, 2002
%and at Strings 2002, Cambridge,   July 15-20,2002} }

\Date{September 2002}

%%%%%%%%%%%%%%%%%%%%%%%%%%%%%%%%%%%%%%%%%%%%%%%%%%%%%%%%%%%%%%%%%%%
\noblackbox
\baselineskip 16pt plus 2pt minus 2pt
%\baselineskip 20pt plus 2pt minus 2pt

%%%%%%%%%%%%%%%%%%%%%%%%%%%%%%%%%%%
\newsec{Introduction}
%%%%%%%%%%%%%%%%%%%

The AdS/CFT duality \mal\ relates the string theory  in \adss 
to $\N=4$  $SU(N)$ SYM  theory in 4-d flat space  with parameters 
${R^4\ov \a'^2}= \l\equiv  g^2_{\rm YM} N $, \ $ g_s = { \l \ov N}$. 
The ``classical'' (tree-level)  string theory limit 
 corresponds to the  `t Hooft limit: $g_s \to 0$, ${R^4\ov \a'^2}= $fixed.
Full understanding of the tree-level   string theory in \adss
would allow one to compute various  ``interpolating functions''
$f_i( \l)$ that enter the observables like 
SYM entropy $S= f_1 (\l)  N^2 V_3 T^3 $,   rectangular Wilson loop 
expectation  value (``quark-antiquark''  potential)
$ \langle W(C)\rangle  \sim \exp [ - f_2(\l) { T \ov L} ]$,  anomalous  dimensions 
of composite operators 
$\Delta - \Delta_0 = f_3 (\l)$ 
 (related to masses of the corresponding string modes), etc. 
While  exact computations of these functions 
 is beyond  the reach at present, 
finding  the leading terms  in them  in  small $\a' \sim \l^{-1/2}$ 
(large $\l$) expansion is already of interest, as  they  provide information 
about strong-coupling behaviour of the SYM theory. 
In  some cases 
%(e.g., in the presence of additional large parameters)
 certain structures  in the  $\a'$   expansion 
of $f_i(\l)$   may  be given exactly 
by the first  few (say, classical and 1-loop) 
 terms in  the  semiclassical expansion.  
This   would then allow the comparison with perturbative  gauge theory results. 
This is indeed  what happens 
in the large R-charge 
example considered in  \rf{\bmn,\gkp} (see also \polk).

Given that the \adss  string action \MET\ 
 is  highly non-linear, 
the semiclassical $\a'$-expansion  near a particular 
string configuration  is the simplest way 
 to explore the string-theory side of the duality. 
Such  expansion was developed  earlier in the open-string (Wilson loop)  sector
\rf{\malrey,\kat,\thei,\dgt}. 
Recently,  similar semiclassical approach  was suggested 
in the closed string sector \rf{\gkp,\frt}, 
where  one 
can relate the  energy of a particular string state to 
the  dimension of the 
corresponding operator in dual gauge theory. 
While the  interest to a particular large $S^5$ angular momentum 
sector of closed string modes  was   drawn  by the observation  \met\ 
that  GS string  in the R-R  plane-wave background \blau\   is exactly solvable, 
one does not actually need to use   the results of \rf{\blau,\met,\mets} 
to arrive at the conclusions of  \bmn:  as explained in 
\rf{\gkp,\frt},   
 all one needs to do is  to expand  the original \adss   action 
\MET\ near a particular  point-like  string classical solution
and compute the 1-loop correction to the energy.
%\foot{In any case, let us  note that recent  surge of activity 
% in the field is  at least to some  extent  due to our improved 
% understanding of the  GS action in R-R backgrounds.}

The semiclassical expansion  approach 
is  more general than the 
one based  on starting directly with the plane-wave  background: \ \ 
(i) it clarifies   the  place  of the  large R-charge sector in the context 
of the standard AdS/CFT correspondence: one concentrates  on
specific  string states  with large $S^5$ angular momentum 
(avoiding   possible ambiguities in 
how one is to take the  Penrose limit  or 
identify  $p^+$ and $p^-$ with $J$  and $\Delta-J$, cf. \rf{\bmn,\nast}); 
the plane-wave (Penrose)  limit is recognized as being  
simply the sigma model 1-loop 
approximation.\foot{This suggests, in particular, 
that it  may be somewhat misleading 
 to  try to find  some  special holography in this case
that would extend to the  full interaction level.}
(ii) it  makes possible    to systematically  extend the 
computation of the energy/dimension  and thus the 
check of the AdS/CFT correspondence 
 beyond the leading (1-loop) order \frt. 
\ \ 
(iii) it  allows one  to investigate 
(by expanding, in  much the same way,  near other  classical solutions) 
 other interesting subsectors  of string 
 states  with large quantum numbers which can be 
again related to particular   operators in   gauge theory.

One may start with any  stable classical string   solution\foot{The classical 
string solutions will  depend only on the string-frame 
metric and (in the extended string case) on the NS-NS $B_{mn}$ field.
However,  the form of the sigma model quantum  corrections
(and thus  a possibility of interpolation  to weakly-coupled gauge theory results) 
will be sensitive to detailed structure  of other background fields 
(dilaton and  R-R field  strengths).} 
carrying     linear or  angular  momentum, or 
simply having  non-zero energy  proportional to the string tension $\sim \sqrt \l$
 times an ``oscillator number'' 
(a ``non-topological soliton''),  
 which is thus large in the $\a' \to 0$ or 
$\l \gg 1$ limit.
For the simplest  point-like string solution, 
one has two ``irreducible'' choices: (i)  massless   geodesic running parallel 
to the boundary of $AdS_5$, and \ \ 
(ii) massless 
geodesic running along big circle of $S^5$ 
(carrying  angular momentum $J$).
In the latter case the 1-loop approximation  is equivalent 
to exact string quantization in the plane-wave (Penrose) limit
of the \adss geometry. 
The basic  extended string  configuration is the 
string rotating  near the center of $AdS_5$
(and carrying spin $S$) \rf{\vega,\gkp}.
% (see also \kar).
One may also  consider  the ``mixed'' $(J,S)$  case  \frt.
Solutions with oscillations 
were considered 
in \rf{\ves,\gkp} and \rf{\rus,\min}
(see also \others\ for other related 
examples in the context  of the semiclassical approach). 
Computing quantum string corrections to the energies
of the  string states  allows one to determine the strong-coupling 
expansions of the anomalous dimensions of the corresponding 
composite operators on the SYM side.

The  study of special sectors  of string states  with large quantum numbers 
is equivalent to semiclassical expansion  since the energy and conserved charges 
then scale as  string tension and thus  are large in the  large $\l$ limit.
 From  general  perspective,  given the non-linearity of the  \adss string action, 
one may try to  do semiclassical  expansions near different points in 
the classical solution space and then try to patch  the resulting expansions  
together. Interpolating  between the  expressions for the string spectra 
obtained near  different expansion  points may lead to a progress  in 
understanding  the  structure of the  string spectrum  in \adss.
An example of such interpolation was described in \frt\ and will be reviewed 
below.

It is useful to recall how similar  correspondence 
between expansions near  different classical solutions 
is achieved in flat space.
The standard  option is to 
 expand near massless point-like  string solution $x_0=x_9= p \tau$. 
Gauging away fluctuations of $x_0+ x_9$, 
  this is equivalent to \lc gauge quantization 
where excited  string modes are identified with
  small fluctuations near  point-like 
vacuum (``supergraviton'') state. 
One finds that the one-loop approximation here is exact and 
$P^+ = E + P_9 = {2 p \ov \a'}, \ \ 
P^- = E - P_9 = { 1 \ov P^+}  ( P^2_\perp +  { 2 \ov \a'} 
\sum^\infty_{n=-\infty} |n| N_n) $.
Alternatively, one may  expand  near the  stationary classical 
solution describing folded  closed string rotating near its center of mass: 
$x_0 = \k \tau, \ \   
x_1 = r(\s) \cos w \tau, \  x_2 = r(\s) \sin w \tau,$
$r(\s) = { \k \ov w}  \sin w \s$, where the 
periodicity in  $\s$ implies quantization condition $w=n=1,2,...$.
The  solution  with $n=1$  has the classical energy and spin related  by 
$E= \sqrt{ {2 \ov \a'} S }$, i.e. it corresponds to  a closed-string 
state on the  leading Regge trajectory in the standard oscillator 
vacuum, $(a_1^\dagger a_{-1}^\dagger)^{S\ov 2} |0\rangle $. 
Expanding near this solution one finds  a tower
  of oscillator string states 
with  given angular momentum. 
The ground  state in this sector (representing the 
unexcited classical solution)  may be described as a coherent state 
of oscillator string  states,  $ |O\rangle _S \sim    e^{ \sqrt S a_1^\dagger}
e^{ \sqrt S a_{-1}^\dagger} |0\rangle  $. 
Similar expansions  may be developed near other oscillating  string solution.
Given that the flat-space string  action is essentially  quadratic in the  conformal gauge 
so that  one knows the general classical string solution  one is able to 
establish correspondence between  expansions near different classical starting points.
Detailed relations  between different expansion points 
in the case of strings in \adss should be of course much more intricate. 

Below  we shall first review the  form of the GS action in \adss 
 in the ``non-conformal'' 
\lc gauge (with the \lc directions parallel to the boundary of $AdS_5$)
and discuss point-like string solutions 
in Poincare and global coordinates (section 2).
We shall then develop semiclassical expansion near  rotating string solutions, 
by considering in turn the $S^5$-boost  $J\not=0$, $AdS_5$ rotation 
 $S\not=0$ and  the mixed $J,S\not=0$ cases (section 3).
In section 4 we shall suggest  an  extension of the semiclassical approach to 
more ``realistic''  near-AdS (near-conformal) cases.

%%%%%%%%%%%%%%%%%%%%%%%%%%%%%%%%%%%
\newsec{Light-cone gauge  GS  action  and point-like classical solutions}
%%%%%%%%%%%%%%%%%%%

Let us start with recalling the form of the quadratic fermionic term in the
type IIB  GS action 
in  \adss 
\eqn\fer{
L_F=
i (\sqrt {- g} g^{ab }\delta^{IJ} -
\ep^{ab } s^{IJ} ) \bar \vt^I \vr_a D_b \vt^J   + O(\theta^4)\ ,
}
where
 $I,J=1,2$,  $s^{IJ}=$diag(1,-1),
$\vr_a \equiv \G_{A} E^{A}_M \del_a X^M, $
 and $E^{A}_M$ is
the
vielbein of the 10-d target space \adss metric
(see \refs{\MET,\dgt} for  details).
In the conformal gauge
$\sqrt {-g} g^{ab} = \eta^{ab}=$diag(-1,1).
The covariant derivative $D_a=\del_a X^M D_M $  is the projection of the
10-d derivative
$D^{IJ}_{ M}= (\del_{ M}
+ \fourth\omega^{AB}_{ M} \Gamma_{AB}) \delta^{IJ}
 - { 1 \ov 8 \cdot 5!}
  F_{ A_1... A_5} \G^{ A_1...A_5} \G_{ M}\epsilon^{IJ}  $.
Since $\theta^I$  are  10-d MW  spinors of the same
chirality   and since $F_5 = R^{-1}(\ep_5 + * \ep_5)$ for the
 \adss background,  $D_a$ 
can be put into  the  following form
\eqn\form{
D_a\theta^I   = (\delta^{IJ} {\D}_a
- { i \ov 2 } R^{-1} \epsilon^{IJ}  \G_* \vr_a ) \vt^J\ ,
\ \ \ \ \ \  \G_* \equiv i \G_{01234} \ , \ \
\G_*^2 =1 \ ,
 }  where
 $ {\D}_a = \del_a
+\fourth \del_a  X^M \omega^{AB}_M\Gamma_{AB
}$. Thus the  action contains a  fermionic 
 ``mass term''  originating  from the R-R coupling \MET.

As usual, the GS   fermionic kinetic term  contains  a potentially problematic 
 factor  $\del X$. 
However,  a   perturbative 
expansion near  a particular $X\not=\const$  string configuration  is well-defined, 
producing (after fixing  an appropriate $\kappa$-symmetry gauge) 
a non-degenerate  fermionic propagator. 

An  alternative  is to use, as in flat space, 
 the \lc gauge \mtt.  
The fermionic $\k$-symmetry gauge $\G^+ \theta^I=0$ 
can be supplemented  by $x^+ = p^+ \tau$ (this  can be justified using, e.g., 
the phase-space approach \mtt).
Here $x^+$ is a \lc direction parallel to the boundary in the Poincare 
coordinates, 
i.e. corresponding  to a \lc  direction in the  gauge theory.
A short-cut way to  arrive at the resulting \lc gauge action is 
to use, instead of the conformal gauge,  the ``diagonal gauge'' \poly\ 
\eqn\dii{
\sqrt {-g} g^{ab} = {\rm diag} (-z^{2},z^{-2}) \ , \ \ \ \ \ \    z=z(\tau,\s)\ ,   }
where $z$ is the ``radial'' coordinate of 
$AdS_5$ in the  Poincare  parametrization, i.e.
($m=0,1,2,3,$ \ $  p=1,...,6 $)
\eqn\asa{
(ds^2)_{_{AdS_5 \times S^5}}= {R^2 \ov z^2} (dx_m dx_m + dz_p dz_p  ) \ , \ \ \ \
\ \ \ \ z^2 = z_p z_p 
\ . } 
In what  follows we shall often set $R=1$. 
The bosonic part of the string action is then  ($\xi^a = (\tau,\s)$)
\eqn\boos{
I = \ha \T \int d^2 \xi \ \big[ \dot x_m \dot x_m  + \dot z_p  \dot z_p 
- { 1 \ov z^4} (  x'_m  x'_m  +  z'_p  z'_p )  \big] \ , 
\ \ \ \ \ \     \T \equiv { \sqrt \l \ov 2 \pi } = { R^2 \ov 2 \pi \a'} \ .  } 
One can then impose  $x^+ = p^+ \tau$ 
(we shall use  decomposition $x^m = (x^+, x^-, x^s),$ $s=1,2$)
 and  derive  the corresponding \lc Hamiltonian that will 
coincide with the one in \mtt. 
Splitting the fermionic components  into 4+4 complex Grassmann variables $\theta_i$, 
$\eta_i$ transforming in fundamental representation of  $SU(4)$,  
the  \lc gauge Lagrangian  can be put into the form \rf{\mtt,\tss} \
(rescaling $\s$ to absorb $p^+$  and  overall effective 
string tension $\T$  factors)
$$
{ L}  =
{  1 \ov 2} \big[ \dot{x}_s \dot x_s   + ( \dot z^p  - {\rm i} 
\eta_i  \rho^{pq}{}^i_j \eta^j  z_q  z^{-2} )^2 
+  {\rm i} (\theta^i \dot{\theta}_i
+\eta^i\dot{\eta}_i - h.c.)  $$  $$ 
-  \ z^{-2} (\eta^2)^2 
- \  \T^2 z^{-4}  (x'_s x'_s  + z'_p  z'_p) \big] 
$$ \eqn\lool{   
- \  \T  \big[\ z^{-3}\eta^i \rho_{ij}^p z_p
(\theta'^j - {\rm i}\sqrt{2} z ^{-1} \eta^j x' )+h.c.\big] \ . 
 }
Here 
$\rho$-matrices  are blocks of $\G$-matrices  and   $x= x_1+ i x_2$.
Like many \lc gauge actions in curved space this action  is not manifestly 2-d Lorentz invariant, but  has a  well-defined  kinetic term. Expanding the 2-d fields 
in Fourier modes in $\s$  one gets a non-linear quantum-mechanical system 
for the infinite number of modes. The theory \lool\ has two obvious limits:
(i)  the ``particle theory limit'' $\T \to \infty$ (i.e. $\l \to \infty$) in which all fields  become independent 
of $\s$ and the spectrum  of the corresponding Hamiltonian 
is the same as the spectrum of the supergravity modes in \adss background \ruslan;
(ii) the ``tensionless string limit'' $\T \to  0 $
(i.e. $\l\to 0$) 
in which all the $\s$-derivative terms  in \lool\ are to be omitted 
but the fields still depend on $\s$ \rf{\mtt,\tss};
in this case,  the presence of the non-linear interaction 
terms involving $z_p$  and  $\eta$-fermions  in \lool\
 implies that the string does not split 
into an infinite collection of decoupled oscillators as that happens 
 in flat space \tss. 
Detailed study of this limit and the spectrum  of the \lc theory \lool\ 
remains an important open problem.

The non-linearity of \lool\  in $z$  (and 
the singularity of the interaction terms near the boundary of $AdS_5$
where $z\to 0$) 
suggests   that to be able to learn more about the quantum 
properties of this theory (e.g., using standard  perturbation theory) 
one needs to expand  near  particular string configurations with 
{\it non-zero}  background values of $z$. 

The simplest classical string configurations 
are point-like null geodesics in \adss  space. 
 They  are easy to find from the action \boos\ in the diagonal gauge.
In general, for  the special class  of string backgrounds 
for which $z(\t,\s) = f(\t) h(\s)$  the 
solutions in the standard conformal  gauge and in the diagonal gauge \dii\ 
are related by the 2-d coordinate transformation:
\eqn\tre{
\tau \to F(\tau) \ , \ \ \  \s \to H(\s) \ , \ \ \ 
\dot F (\t) = f^2(\t) \ ,\ \ \ \   H'(\s) = h^2(\s)  \ , \ \ \ 
z(\t,\s) = f(\t) h(\s) \ . } 
The  $\s$-independent (point-like) solutions  of equations of motion 
 plus constraints corresponding to  
 \boos,\dii\ 
are the  same straight null lines as in flat space
($a_m,p_m, u_p,v_p = \const$)
\eqn\mull{
x_m = a_m + p_m \tau \ , \ \ \ \ z_p = u_p + v_p \tau  \ ,\ \ \ \ \ \ 
p_m p_m + v_p v_p =0  \   .}
Transforming this solution to the conformal gauge  using  \tre\ 
one finds the general form of point-like string solution in \adss 
which was previously given in \kat\ 
\eqn\oldi{  
x_m = a_m + p_m q(\tau) \ , \ \ \  \ \ \ \ \ \   \ z_p = u_p + v_p q(\tau)
  \ ,  }
\eqn\ght{
 q(\t) = { \om \ov v^2} \tan (\om \tau)  - { u \cdot v  \ov v^2}  \ ,\  \ \ \ \ \ 
\om^2 \equiv u^2 v^2 - ( u \cdot v)^2    \ . }
It is easy to  indentify the  two    non-trivial cases:

(I) Null geodesic parallel to the boundary  of $AdS_5$:
\ \ \   If $p_m p_m \equiv -p_0^2 + p_i^2 =  - v_m v_m =0$, then  $v_m=0$
and the solution is the null geodesic parallel   to the boundary
located at a distance $z$ from the boundary ($z_p= u_p =$const). 
We can choose  $p_m$ and $u_p$ so that 
the solution  is given by\foot{It thus has the same  form in the 
 conformal   and  diagonal gauges.}
\eqn\gre{
x_0=x_3 =\ p\ \tau \ ,\ \ \ \ \ \ \  z_1 = u = \const \ , \ \ \  \ \ \ \ 
x_1,x_2, z_2,..., z_6=0 \ . }   
The expansion near this classical solution (with fluctuations 
of $x^+= x_0 + x_3$ gauged away) is equivalent to  the 
expansion of the \lc action \lool\ near the constant  $z_p$ point. 
Here the energy is $E=P_0=P_3 = \sqrt \l p  \gg 1 $. 
It is easy to see from \lool\ that here (in contrast to the case of the  $S^5$-geodesic  
discussed below) the  resulting quadratic fluctuation action is  
trivial, i.e. is  the same as in flat  space, giving the 1-loop correction to the energy 
$E_1  = { 1 \ov 2p} \sum |n| N_n$. 
However, the 2-loop corrections to the energies of strings states will be  non-trivial 
($E_2 = { 1 \ov \sqrt \l}  F(p,n)$) and  would be 
 interesting to compute   explicitly.

(II)  Null geodesic transverse to the boundary of $AdS_5$
(along the big circle of $S^5$). \ \ \  \ 
If $p_m p_m =  - v_m v_m $  is non-zero, then 
the  translational invariance in $x_m$ and $SO(1,3) \times SO(6)$ rotational
 symmetry allows us to  bring the corresponding solution in the  diagonal gauge  \mull\
to the form 
\eqn\sdi{
x_0= p \t \  , \ \ \ \ \ \ z_1 =  u= \const\   , \ \ \ \ z_2=p \t  \ ,   }  
where we  also used the  translational invariance  in $\tau$ to set $u_2=0$.
This is a straight-line geodesic  in the $(z_1,z_2)$ plane   parallel 
to the $z_2$-axis. 
Note that the expansion near this solution  can  be again  described by the \lc 
action \lool\  (the \lc gauge choice means only that the ``boundary'' \lc  coordinate
 $x^+$ does 
not contain  quantum fluctuations).

This second solution \sdi\ 
corresponding to  a straight motion  of a particle  in the plane $(z_1,z_2)$ 
is carrying a   non-zero angular momentum since  $z_1\not=0$. It 
can also be viewed  as a null geodesic in \adss running 
  along the big circle of $S^5$. To show this 
let  us first transform \sdi\ to the conformal gauge, i.e. write down the solution 
\oldi\ with 
$u_p= (u,0,0,0,0,0), \ v_p=(0,p,0,0,0,0), \ u \cdot v=0, \ a_0=0, \ p_0 =p$:
\eqn\sdip{
x_0= p \tan \om  \t \  , \ \ \ \ \ \ z_1 = u \ ,\  \  
\ \ \ z_2 = p \tan ( \om  \t )  \  ,  \ \ \ \  \ \ \ \   \om= u p  \ .  }  
For  simplicity let us  also rescale $x_m$ and $z_p$ 
by $1/p$ and choose $ u=p$.  Then  \sdip\  can be written as 
\eqn\sdipe{
x_0=  \tan  t    \  , \ \ \ \ \ \ 
z_1 =   1 = z   \cos \vp    \ , \ \ \ 
z_2 = \tan t =     z   \sin \vp  \ ,\ }
where 
\eqn\zes{  t= \om \tau\ , \ \ \ \  \vp = \om \tau  \ ,\ \ \ \ \  \   
 z=  { 1 \ov \cos \om \tau}   \ . }
Here $\vp$ is, in fact, the  angle of large  circle of $S^5$ 
and $t$ is the global time  coordinate of $AdS_5$. 
Indeed,  in  global coodinates 
\eqn\add{
ds^2_{_{AdS_5}}
= - \cosh^2 \r \ dt^2 +  d\r^2 + \sinh^2\r \ d\Om_3 \ , }
$$
d\Om_3
= d \b_1^2 + \cos^2 \b_1 ( \ d\b_2^2 +
 \cos^2 \b_2  \ d \beta_3^2)   \ ,
% \ \ \ \ \   \beta_3\equiv \p \ , 
 $$
while the  angle   $\vp$ 
of $S^5$  related to $z_p$ coordinates by 
$dz_p dz_p = dz^2 + z^2 d\vp^2  + dz_n dz_n$.

In  general, the 
 transformation between the Poincare and the global coordinates of $AdS_5$ 
can be  done as follows (we set the radius of $AdS_5$ 
 to be 1 and use the Minkowski signature):
\eqn\tran{
X_0={x_0\ov z} = \cosh \r \ \sin t \ , \ \ \
X_5= { 1 \ov 2 z} ( 1 + z^2 - x^2_0 + x_i^2) =  \cosh \r \ \cos t \ , }
\eqn\med{
X_i= {x_i\ov z} = n_i \sinh \r  \ , \ \ \  
X_4={ 1 \ov 2 z} ( -1 + z^2 - x^2_0 + x_i^2) =  n_4 \sinh\r \ , 
\ \ n^2_i + n^2_4 =1 \ ,   }
\eqn\sed{ \tan t = { 2 x_0 \ov 1 + z^2  - x^2_0 + x_i^2 }   \ , \ \ \ \ \ \ \ \ 
z^{-1} = \cosh \r \ \cos t - n_4 \sinh \r  \ . } 
Here $X_0,X_i,X_4$ ($i=1,2,3$) 
are the coordinates of  $R^{2,4}$:
the $AdS_5$ metric is induced from the flat $R^{2,4}$ one by the  embedding (see, e.g., 
\mal)
$X_0^2 + X_5^2 - X_i^2 - X^2_4 =1$. 
The unit vector 
 $n_k$ ($k=1,2,3,4$), \     $n^2_i + n^4_4=1$, \ 
 parametrizes the 3-sphere: \ $dn_k dn_k = d\Omega_3$. 
The obvious point-like solution in global 
coordinates \add, i.e. $t= \om \tau, \ \r=0, \  \vp= \om \tau $
 (with all other angles being trivial), 
then becomes  equivalent  to 
\zes.

Expanding the covariant \adss string action (in the conformal gauge 
   or in diagonal gauge  \boos) 
 near the above solution one reproduces (see \frt\ and below), 
in the 1-loop approximation, the same spectrum as found in
 the case of the plane-wave background \rf{\met,\mets,\bmn}.\foot{The 
1-loop approximation (i.e. taking  $\a'\to 0$ and rescaling the coordinates) 
is related to the  strict Penrose limit \sfets.}

%%%%%%%%%%%%%%%%%%%%%%%%%%%%%%%%%%%%%%%%%
\newsec{Semiclassical expansion near  rotating string  solutions}
%%%%%%%%%%%%%%%%%%%%%%%%%%%%%%%%%%%%%%%

%%%%%%%%%%%%%%%%%%%%%%%%%%%%%%%%%%%%%%%%%%%%%%%%%%%%%%%%%%%
\subsec{\bf Classical energy and classical solutions}
%%%%%%%%%%%%%%%%%%%%%%%%%%%%%%%%%%%%%%%%%%%%%%%%%%%%%%%

%Before describing the  string  solutions we will be interested in 
Let us  first  supplement the transformation between 
the Poincare and global coordinates of  $AdS_5$ 
\tran--\sed\   by the relation between the  corresponding  energies
(corresponding to $t$ and $x_0$ time coordinates). 
The energy in global coordinates expressed in terms of the   Poincare coordinates
$x_m,z_p$ in \asa\ 
as functions of  both  $\tau$ and $\s$  is\foot{Note
 that in general $E$ may not be conserved  if solutions do not
decay fast enough in $\s$:
the translation invariance in $t$ implies that 
$ \del_a E^a=0, \   E^a = {\del L \ov\del  \del_a t } , 
$ so  that one needs  $\int d \s \ {\del L \ov\del  \del_\s t } =0$
for the energy to be conserved.} 
$$E= \sql \E =  \sql\int {d \s\ov 2\pi}  \E_d = 
    \sql\int {d \s\ov 2\pi}  { \rm cosh}^2\r\  \dot t 
$$
\eqn\eee{
 =   \ha   \sql\int {d \s\ov 2\pi}  \big[ (1 + z^2  + x^2 ) \PP_0 - 2 x_0 \DD \big
] \ ,  
}
\eqn\tyi{    \PP_0 =   {1 \ov z^2} \dot x_0 \ ,  \ \ \ \ \  \ \ \ \ 
\DD=   {1 \ov 2z^2} { \del \ov \del \tau} ( z^2  + x^2 )   \ , \ \ \ \ \ \ 
  x^2= - x_0^2 + x_i^2\  .  } 
Here  $\PP_0$ is the  energy density corresponding to translations in $x_0$ and 
$\DD$ is  the dilatation  ($z\to  k  z, \ x_m \to  k  x_m $) 
charge density. 
One can show that  $E$  is  related to the 
superconformal generators  in the Poincare patch
 as  follows\foot{At the boundary of $AdS_5$  (i.e. $z=0$)
the standard superconformal generators are 
$\D= - i x^m \del_m, \ \ \P_m = - i \del_m , \ \ 
\K_m = - 2 x_m \D + x^2 \P_m  = i ( 2 x_m x_n - x^2 \eta_{mn} ) \del_n , 
$ $ \ x^2 = x_m x_m = - x^2_0+ x_i^2 , \ $ and so 
$[\D, \P_0] = i \P_0,  \ [\D, \K_0] = -i \K_0, 
\ [\P_0, \K_0] = -2i \D , \ $ etc.
The above generators acting on the $AdS_5$
 coordinates are extensions of these to the bulk.}
\eqn\ses{  E= \ha ( P_0 + K_0) = \ha       \sql\int {d \s\ov 2\pi}        ( \PP_0 + \KK_0) 
  \ , \ \ \ \ \  \ \ \ 
\KK_0 = ( z^2 +  x^2) \PP_0   - 2 x_0 \DD   \ . }
We can also express the energy density in \eee\  as 
\eqn\sre{
\E_d= \ha { (1 + z^2 + x^2)^2 \ov z^2 } {\del \ov \del \t}(  { x_0 \ov
 1 + z^2 + x^2})  \ . }
For the solutions  discussed below  one has the following relation 
 \eqn\spe{
 z^2 - x^2_0 + x^2_i = 1       \ . }
Thus $\DD=0$ (see \tyi)  and  
    the energy $E$ in global coordinates \sre\  coincides with the energy $P_0$
 in the Poincare coordinates:
\eqn\dre{
E= P_0  = K_0 =   \sql\int {d \s\ov 2\pi}   \ { 1 \ov z^{2}} \dot x_0  
\ , \ \ \ \ \ 
\ \ \ \ \   \DD=0 \ . }

Next,  let us    summarize the (conformal-gauge) form of the 
simplest classical string solutions  in \adss 
written  in global \add\ and Poincare \asa\ coordinates of $AdS_5$
($G$ will stand for the  global and $P$ for the  Poincare coordinate form).
As was already  discussed above, for  the 
point-like string  rotating in $R^6$ (or, equivalently,   boosted 
along a big circle in $S^5$)
\eqn\poi{
G: \ \ \ \   t= \n \t \ , \ \ \ \ \ \ \  \vp = \n \t  \ , }
\eqn\poo{ P: \ \ \   x_0 = \tan t \ , \ \ \  z= { 1 \ov \cos t}  \ , \ \ \ \ \ 
\vp =  t = \n \t \ . }
The spinning string in $AdS_5$ is described by (the angle $\p$ is $\beta_3$ in \add) \rf{\vega,\gkp}
\eqn\spin{
G: \ \ \ \   t= \k \t \ , \ \ \ \p = w \t \ , \ \ \r= \r(\s) \ , \ \ 
 \ \  \r'^2 = \k^2 \cosh^2 \r - w^2 \sinh^2 \r   \ , }
or, equivalently, by 
\eqn\pin{  P: \ \ \  
 x_0 = \tan t \ , \ \ \  z= { 1 \ov \cos t 
\ \cosh \r} \ , \ \ \
 x_1 = r \cos \p \ , \ \  x_2 = r \sin \p \ , 
 \  \  r\equiv  { \tanh \r\ov  \cos t} \ , }
where $t=t(\t)$, \ $\p=\p(\tau)$, and $\r=\r(\s)$ are given by 
\spin. 
Here  $\r$ changes from 0 to its maximal value $\r_0=$Arctanh${\k\ov w} $
and 
the parameter $w$  is a function of $\k$.
In Poincare coordinates the string moves towards 
the  horizon (center of AdS), rotating and stretching. 
More general solution can be obtained  by combining the  rotation in $AdS_5$ 
with the  boost in $S^5$ \frt\ (see section 3.4 below).  

One may also consider  string spinning in $S^5$ \gkp, for which 
$t= \k \t  ,  \ \vp = \n \t  ,  \ \psi= \psi(\s) $
($\psi$ is an angle of $S^5$ along which the string is stretched, $ds^2_{S^5} = 
d \psi^2 + \sin^2 \psi\ d \vp^2 + ...$), 
as well as a more general solution interpolating  between 
this and the above two, i.e.  string spinning in $AdS_5$ as well as in  $S_5$
\rus. 

Let us mention also some of the   oscillating string solutions 
\rf{\ves,\gkp,\min}. 
For  the string oscillating in $AdS_5$ 
 \eqn\drew{ 
G: \ \ \ \   t=  t(\t) \ , \ \ \ \p = w \s \ , \ \ \r= \r(\t) \ , \ \ 
 \ \    \dot t = {\k \ov \cosh^2 \r} \ , \ \ \
          \dot \r^2 = {\k^2 \ov \cosh^2 \r}  - w^2 \sinh^2\r  \ . } 
The form of this solution  in 
 Poincare coordinates  is similar to the one   in  the rotation case \pin\ 
$$ P: \ \ \   x_0 = x_0 (\t) \ , \ \ \  x_1 = r \cos \p \ , \ \
x_2 = r \sin \p \ ,   $$ 
\eqn\wed{  
{ r \ov z}=  \sinh \r \ , \ \   { x_0 \ov z} = \cosh \r \sin t 
\ , \ \ z= { 1 \ov \cosh \r \ \cos t } \ , \ \ x_0 =\tan t  \ . } 
For the string oscillating  in $S^5$ one has 
$t= \k \t  ,  \ \vp = w \s  ,  \ \psi= \psi(\t) .$
There are also more general solutions incorporating the above as special cases
\rru. 

The form of a 
 classical  solution cannot  depend on the value of the string tension,  i.e. on $\sql$, which appears as  a factor in front 
the string  action $I= {\sql \ov 4 \pi} \int d^2 \xi \ G_{MN} (x) \del_a x^M \del^a  x^N $. Thus the classical energy 
can be written as  $E= \sql \E(\cN)$, where $\cN$ stands for all 
constant parameters (like $\nu,w,\k$ in the above expressions \poi--\wed)
that enter the classical solution. These parameters should be 
 fixed in the standard sigma model loop ($1 \ov \sql $) expansion.
%i.e.  $\E$ cannot depend on $\sql$. 
However,  some of them may be quantized in the full quantum theory, 
i.e. $\sql \cN = \rN$=integer
(being related to canonical momenta the  quantized parameters should contain 
a factor of string tension). 
For example, the  global charges  like  the 
$S^5$ and $AdS_5$ angular momentum  components 
$J= \sql \nu, \ \ S= \sql \S (\k)$, etc.,   will  take  integer values.
In addition, for the oscillating  solutions there will be 
an integer  ``oscillation number'' parameter 
$\rN_{osc}= \sql \cN_{osc} \gg 1 $:
as usual for stationary ``non-topological'' soliton-type  solutions
 there  will be  a  semiclassical path integral phase  quantization 
 condition (see  \rf{\dash,\ves}).
Expressed in terms of these quantized parameters  the  classical 
energy will look like 
\eqn\cla{E= \sql\  \E\big( {J\ov \sql}, {S\ov \sql}, {\rN_{osc}\ov \sql}\big)  \ .}
According to the  AdS/CFT duality 
\gkp, the energy in global coordinates
  should give the expression  for the strong-coupling 
limit of the (canonical+anomalous) dimension  of the 
corresponding  SYM operator. 
The string sigma model corrections to \cla\  discussed below 
  will represent subleading strong-coupling expansion  terms in the anomalous dimension.
%\foot{In the planar limit only, as we will not consider 
%string loop corrections.} 

As was  observed in \ves,  in the limit of 
 large values of $\cN$ (and thus of  the quantum 
numbers $\rN\gg 1$)  the classical energy  of a string  solution in any $AdS_p$  space 
goes as {\it  linear} function of $\rN$, i.e.  
$E \sim \rN $. Here 
 $\rN$ can be   an angular momentum 
 or an  oscillation number or a combination of the two.
 This is to be 
compared to $E \sim \sqrt \rN$ in  the  flat space case.
 This remarkable linear behaviour 
(seen explicitly on specific examples of solutions 
 in \rf{\vega,\gkp,\frt,\rus,\min})
is  a consequence of  the curvature of the 
$AdS_p$ 
and is perfectly  consistent with the  AdS/CFT duality:  the large $\rN$ 
expression for 
$E$ (i.e. of  the full dimension) should start with 
 canonical  dimension of  the corresponding gauge-theory  operator.

Below we shall discuss the classical 
 relation \cla\  
and   quantum string sigma model corrections to it
on specific examples. 
Our aim will be to try to draw some general conclusions 
about the expression for the quantum energy 
$E$  on the string side of the AdS/CFT duality.

% \eqn\oss{
%G: \ \ \ \   t=  t(\t) \ , \ \ \ \vp = w \s \ , \ \ \r= \r(\t) \ , \ \ 
% \ \  \dot t = {\k \ov \cosh^2 \r} \ , \ \ \
%    \dot \r^2 = {\k^2 \ov \cosh^2 \r}  - w^2   \ , }  while for

\bigskip

%%%%%%%%%%%%%%%%%%%%%%%%%%%%%%%%%%%%%%%%%%%%%%%%%%%
\subsec{\bf Quantum corrections  to  point-like string  rotating in $S^5$}
%%%%%%%%%%%%%%%%%%%%%%%%%%%%%%%%%%%%%%%%%%%%%%%%%%%%%%%%%%%%

Let us now consider  the  quantum  corrections to the energy of 
the point-like $S^5$-rotating solution \poi.
Its classical energy and angular momentum 
$E= \sqrt \l \n $ and $ \ J= \sqrt \l \n$ 
are proportional  to the   parameter $\nu$ 
which  should be { fixed} 
  in the semiclassical expansion in powers of the  inverse string 
tension $1 \ov \sqrt \l$. Then  $J \gg 1 $  in the 
large  $\sqrt \l$ limit. 

 The general strategy is as follows.
One is supposed to expand 
the string action near  a given  classical solution  and then compute corrections 
in  sigma model and string perturbation theory. 
If  one is interested in quantum  corrections to the energies 
of string states in global coordinates in $AdS_5$ 
%(anomalous dimensions of the corresponding gauge-theory operators)
 one  should consider string theory on the cylinder\foot{For an alternative 
description of the semiclassical approximation for the  2-point function
of vertex operators  in Poincare coordinates see \poll.} 
(or cylinder with  handles attached  if 
 one goes beyond tree level in string coupling)
 and  compute the  expectation value of the energy operator 
$\langle  \Psi| \hat E | \Psi \rangle$ 
 in the sector of states  with 
 quantized  angular momentum  \frt:
$\langle  \Psi| \hat J | \Psi \rangle  = J = \sqrt \l \nu $= integer. 
The  parameters of the string perturbation theory 
are $ { \a' \ov R^2} = { 1 \ov \sqrt \l} \ll 1$, 
$\nu = { J \ov \sqrt \l} $=fixed    and $g_s \ll 1$. 
They can be  expressed in terms of the 
parameters  $\l'$ and $g_2$  \rf{\bmn,\plef,\gro}  that 
enter the SYM perturbation theory for the anomalous dimensions 
of operators with large R-charge:
\eqn\dedf{
\l'\equiv {\l \ov J^2} = { 1 \ov \nu^2}  \ , \  \ \  {\rm i.e.}   \ \  \ \ \
  { 1 \ov \sqrt \l}  = {1 \ov J} { 1 \ov \sqrt {\l'} } \ , \ \ \ \ \ }
\eqn\dedj{
g_2 \equiv {J^2  \ov N} \ , \  \ \  {\rm i.e.}   \ \  \ \ \ 
 g_s = g^2_{\rm YM} = { \l \ov N} = \l' { J^2 \ov N} = \l' g_2  \ . }
Then the semiclassical string limit 
 $\n$=fixed, $\l \to \infty  $, $g_s \to 0$ 
(i.e. the sigma model
  1-loop approximation in  string theory on the cylinder)
is equivalent to $\l'$=fixed, $J\to \infty$, $g_2 \to 0$. 
Choosing $\nu$ to be large  corresponds to $\l' < 1$. 
In general, the expression  for the energy of string states 
computed in both sigma model and string loop perturbation theory will be 
\eqn\sert{
E= E({ 1 \ov \n^2}, { 1 \ov \sqrt \l}, g_s) 
= E(\l', { 1 \ov  J \sqrt {\l'}}, \l' g_2)
%= \sum_{l=0}^\infty  c_l (\nu)  ({ 1 \ov \sqrt \l})^{l-1}  + O(g_s^2)  
\ . } 
Again,  sending $g_2$ to zero and then $J$ to infinity for fixed $\l'$ 
on the SYM side is thus equivalent  to  the sigma model 
1-loop approximation  in tree-level   string theory. 
The energy of  a string state with  some quantum  number $n=\{n_i\}$
(i.e. the expectation value of the quantum space-time energy operator 
$E=\langle  \Psi_n | \hat E | \Psi_n \rangle $) may be written as 
\eqn\see{
E= \sum_{l=0}^\infty  E_l   + O(g_s^2) \ , \ \ \ \ \ \ \ \ \ \ \
E_l = { 1 \ov (\sqrt \l)^{l-1}}  \E_l (\nu; n)   \ ,   }
where $l=0,1,2,...$ is the  sigma model loop order.
Residual  supersymmetry preserved by the classical solution 
implies that $\E_l$ with $l >   0 $ should vanish for the ground
 state (and all other supergravity modes in the spectrum). 

%Let us first discuss the 1-loop ($l=1$) approximation and then comment 
%on  the  form of the higher-order terms. 

Expanding the  bosonic part of the \adss string action near the classical solution
\poi\ or \poo\ one finds the following  quadratic 
term in the Lagrangian  for  the  fluctuation fields \rf{\gkp,\frt}
\eqn\fluu{
L_{2B }= - (\del \td t)^2  + (\del \td \vp)^2 
+ (\del  \td \xi_k)^2  +  (\del  \td\psi_k)^2  + \nu^2 ( \td \xi_k^2   +  \td\psi_k^2 ) \ .  }
Here $\xi_k$ and $\psi_k$ ($k=1,2,3,4$) are 4+4  fluctuations in 
other directions of $AdS_5$ and $S^5$ rescaled by $(\sql)^{-1/2}$.
 The mass terms originate from the 
curvature of the two  spaces. The next  term in the expansion 
of the sigma model Lagrangian 
is a quartic interaction term  which  has 
 the following symbolic structure \frt
\eqn\flu{
L_{4B} \sim  { 1 \ov \sql} \big[
-  \td\xi^2 (\del \td t)^2  +  \td\psi^2 (\del \td \vp)^2 
+ \td \xi^2 (\del  \td \xi)^2  +   \td\psi^2   (\del \td  \psi)^2  
+ \nu^2 (  \td\xi^4   +  \td\psi^4 ) \big] \ ,  }
so that the   parameter $\nu$ of the classical solution 
determines the masses as well as the 
potential terms for the 8 bosonic fluctuation fields.
More generally, the  expansion of the bosonic part of the action 
has the structure (here $\xi$ stands for all 8 dynamical 
fluctuation  fields and
in contrast to \fluu,\flu\  we do not rescale $\xi$ by $(\sql)^{-1/2}$)
\eqn\fla{
L_B  \sim  \sql \bigg[
(\del  \xi)^2  + \nu^2 \xi^2   + \sum_{m=1}^\infty  
\big[ c_m \xi^{2m} (\del \xi)^2 + b_m \nu^2  \xi^{2m+2}   \big]  \bigg]\   . } 
The form  of the quadratic fermionic term is easily found from 
the general expression \fer,\form.\foot{Here 
$\rho_0 = \G_M \del_0 X^M = \nu (\G_0 + \G_\vp), \ \ 
D_a^{IJ} = \delta^{IJ} \del_a  - { i \ov 2} \ep^{IJ} \G_* \rho_a \ , \
\G_* = i \G_{1234} $.} 
Choosing the natural $\k$-symmetry gauge  (which is essentially ``imposed'' on us 
by the choice of the background) $\G^+ \theta^I=0, \ \ \G^{\pm} = \mp \G_0 + \G_\vp$, 
we get \frt
\eqn\swq{
L_{2F}  = - i \nu ( \bar \theta^1 \G^-  \del_+ \theta^1 + \bar \theta^2 \G^- \del_- \theta^2
- 2 \nu \thet^1 \G^- \Pi \thet^2)  \ , }
where $ \Pi \equiv i \G_* \G_0 = \G_{1234}, \ \  \Pi^2=I$, 
and the mass term originated from the R-R coupling. 
Redefining the spinors,  we end up with an action that can be interpreted 
as  describing  4+4    massive 
2-d fermions (with Dirac-type  mass terms, i.e. $ S_R \del_+ S_R + 
 S_L \del_+ S_L  \pm \nu  S_L S_R $). 
Higher-order interacting fermionic terms that accompany \flu\ 
have the following structure 
\eqn\hii{
L_{F} \sim   {1 \ov (\sql)^{1/2} } \big[ \bar  \theta\del \theta ( \del  \td \xi  +  \del  \td\psi) 
+  \nu \bar \thet \thet ( \del  \td \xi  +  \del  \td\psi) \big]
+  {1 \ov \sql} \big[  \bar \thet \thet  ( (\del  \td\xi)^2  + (\del  \td
\psi)^2 ) + 
\nu^2  \bar \thet \thet \bar \thet \thet \big] + ...  \ . } 
The quadratic part of the expanded \adss 
 action \fluu,\swq\ is the same as the full 
GS action for  the plane-wave R-R background of \blau\ 
that was  found in \met. 
This is not surprising being a   consequence of the fact 
that the plane-wave  background is 
 the Penrose limit 
of the \adss near the same null geodesic \rf{\blau,\bmn}.
The parameter $\nu$  can be identified 
with the product of $\a' p^+$ in the \lc 
gauge condition $x^+ \equiv t + \vp = \a' p^+ \tau$ 
 in \rf{\met,\mets} and a scale  of the plane-wave  background. 

%Returning to the 1-loop  approximation, 

To find the quantum correction to the energy  one  may use  the constraints 
to eliminate $\td x^- = \td t - \td \vp$ or, equivalently,  impose the ``light-cone''
gauge $\td x^+=0$. Since  for classical trajectory $t = \nu \tau$, 
the 10-d energy and  the 2-d energy  are directly related (as in the 
familiar case of the static gauge), 
with the proportionality 
factor $1\ov \nu$. The 
quantum correction to the energy  
is then determined by the expectation value of the 2-d Hamiltonian 
 \frt\ 
\eqn\asd{ E-J =   { 1 \ov \nu} \int^{2\pi}_0 { d\s \ov 2 \pi}\   \langle \Psi| 
{\cal H}_{\rm 2d} (\td  \xi,\td \psi,\thet) | \Psi  \rangle   \ . } 
Here ${\cal H}_{\rm 2d}$  corresponds to the
above action \fla,\swq,\hii\ 
 describing   4+4 massive bosonic and 4+4 massive fermionic 
``transverse'' fluctuation modes and their interactions 
as prescribed by the full 
\adss GS action expanded near the classical trajectory. 

Ignoring the interactions, i.e. omitting the 2-loop $O({1\ov \sql})$  and higher 
 corrections to $E$, 
one finds the same expression as in  \rf{\met,\bmn,\mets}
\eqn\rell{
E= \sql \nu + { 1 \ov \nu} \sum^\infty_{n=-\infty} \sqrt{n^2 + \nu^2}  \ N_n 
= \sql \nu +  \sum^\infty_{n=-\infty} \sqrt{1 + {1 \ov \nu^2} n^2 }  \ N_n  \  . }
Expressed in terms of $J$ and $\l$, the classical plus 1-loop correction to the energy is thus 
\eqn\relol{
E=J +   \sum^\infty_{n=-\infty} \sqrt{1 + {\l \ov J^2}n^2 }  \ N_n= 
 J +   \sum^\infty_{n=-\infty} \sqrt{1 + \l' n^2 }  \ N_n \  . }
The analyticity of the latter square root 
 expression in $\l'$  suggests  the possibility of direct comparison 
with the  SYM perturbation theory expression 
 for the corresponding anomalous dimensions obtained  in the limit
 $J \to \infty, \ \l'= $fixed   \rf{\bmn,\gro};  indeed,  one finds the 
exact agreement  with the square root correction  in \relol\  \zan. 

Why this happens is clear from the structure of the string 
perturbative expansion: higher-loop  sigma model corrections are
 suppressed by powers of $1 \ov \sql$
and thus by powers of $1\ov J$  (for fixed $\l'$).
 Still, there is 
a small miracle in  that the string  expression \relol\  has regular expansion in power
series in $\l'$ -- otherwise, one would need to do a non-perturbative
 computation 
on the SYM  side to be able to compare to the above string result. 
One may wonder if the same  analyticity property holds also at higher orders in 
sigma-model loop expansion, i.e. at subleading orders in $1\ov J$ expansion, 
thus 
allowing one  to hope  extend the comparison  with simply 
 perturbative SYM theory beyond $J=\infty$ limit. 

Indeed, it  is possible to argue  that all higher-order terms in \see\ 
should have 
%the following structure (note that $E_l = { 1 \ov \nu} (E_{2d})_l$, cf. \asd)
 the  structure
\eqn\coor{
E_l=
{1 \ov \nu} (E_{\rm 2d})_l =
{1 \ov (\sqrt \l\ \nu )^{l-1}} F_l ( {1\ov \nu^2}; n)  \ , \ \ \ \ \ \ 
  F_l =  c_l (n) +  { 1 \ov \nu^2} d_l (n) +  O({ 1 \ov \nu^4})  \ .  } 
The expanded  GS action defines  an UV finite   massive 
2-d QFT on a cylinder, with interaction vertices containing 
only two derivatives or two powers of $\nu$
 (cf. \fla,\hii).
Thus the energy   of a particular
oscillator string  state  with quantum  numbers
$n$ should have a regular inverse-mass 
 ($1\ov  \nu$) 
expansion, i.e. each $l >   1$ correction $E_l$ in \see\ 
 should vanish in the limit 
of $\nu\to \infty$ (for fixed $\l$).  This implies \coor\foot{It may be possible
also  to show that in general $c_l=0$.}
which may be written also as 
\eqn\coor{
E_l=
 { 1 \ov J^{l-1}}  F_l ( {\l \ov J^2}; n) = { 1 \ov J^{l-1}}  F_l ( \l'; n)
\ , } 
with  $F_l$ having  a regular  power series expansion for  small  $\l'$.
As a result,  in the large $J$  sector, 
one should be  able to re-interpret the string $\a'\sim {1\ov \sqrt \l}$
expansion  as an expansion  in positive  powers
of $\l'={\l \ov J^2}= { 1 \ov \nu^2}$. 

This implies  that the 
  comparison between the string theory and the 
 SYM theory  can be extended to 
subleading orders in $1\ov J$  without need to go 
beyond   perturbation theory in $\l'$ on the SYM side.  
The  leading 2-loop correction to the energy of excited string states
(like $a^\dagger_1 a^\dagger_{-1} |0>$)
has the  following structure \rf{\frt,\tsett,\parn}
\eqn\reff{
E_2 =  {1 \ov \sqrt \l\ \nu } F_2 ( {1\ov \nu^2}; n)  
=   { 1 \ov J}  F_2 ({\l \ov J^2} ; n) = { \l \ov J^3} d_2 (n) + ... 
\  . } 
As follows from \coor, the $ { \l \ov J^3} $ term in the energy
(anomalous dimension) may appear only from the 2-loop string correction
(i.e. there is  no non-trivial function of $\sql$ multiplying $1\ov J^3$ term).
Therefore  that the  coefficient $d_2$ in \reff\  should be 
reproduced by the leading  perturbative (single  YM interaction vertex insertion) 
 computation on the  SYM side (see   \parn). 
It would be interesting to find the  complete string-theory 
expression for the 2-loop  function  $F_2$ in \reff\foot{For that one needs 
to expand the Hamiltonian
(see \asd, \hii)  \ $\cH_{\rm 2d}=\cH_0 + { 1 \ov (\sql)^{1/2}} \cH_1 + 
 { 1 \ov \sql} \cH_2 +... $
and the quantum state $|\Psi\rangle = |\Psi_0\rangle  +   
{ 1 \ov (\sql)^{1/2}} |\Psi_1\rangle  + ... $, 
and to take into account that 
 both $ \langle \Psi_0| \cH_0 |\Psi_0 \rangle$ and 
 $ \langle \Psi_0| \cH_1 |\Psi_1 \rangle$  may  contribute to the 
order ${1 \ov \sql}$ correction.}
  as well as 
to  extend  the gauge-theory computation of \zan\  to subleading order in $1\ov J$. 
A complication in checking the AdS/CFT duality 
at subleading orders in $1\ov J$  is that one needs also to modify \parn\ 
the  definition \bmn\ of the composite SYM operators corresponding 
to the string modes. 

\bigskip 

%%%%%%%%%%%%%%%%%%%%%%%%%%%%%%%%%%%%%%%%%%%%%%%%%%%
\subsec{\bf Quantum corrections  to    string  rotating in $AdS_5$}
%%%%%%%%%%%%%%%%%%%%%%%%%%%%%%%%%%%%%%%%%%%%%%%%%%%%%%%%%%%%

Let us now consider the spinning string solution \spin\ \rf{\vega,\gkp}. 
Here the classical energy (which is the same  in global or Poincare coordinates, 
see \dre)  $E= E_0= \sql \E (\k,w)$  and the spin  $S= \sql \S(\k,w)$
depend on the classical parameters $\k,w$. The latter 
  are related by the $\s$-periodicity condition, $w=w(\k)$, so that 
one can express the energy in terms of spin, 
 $\E=\E(\S)$  or    $E=E(S,\sql)$.
One may  view $\S$ as the basic   parameter of the classical 
solution which  is fixed in the semiclassical  expansion  in $1 \ov \sql$.

 In the  ``short string''  ($\S \ll 1$) 
limit
%\foot{The maximal length of the string is determined by $\coth \r_0 = { w \ov \k}$.}
one finds that $\E \approx \sqrt{ 2 \S} $, 
i.e. 
 $E \approx \sqrt { 2 \sql S}$; this  is the same 
linear Regge trajectory relation  as in flat space
(not surprising,  since the short string is located at  the center of $AdS_5$ which
is approximately flat).  For ``long string'' ($\S \gg 1$) 
one finds that  $\E \approx \S$, i.e. $E \approx S$ \vega, 
in agrement with general behavior of $E$ in AdS space 
discussed in section 3.1. 
%This change of behaviour  of $E(S)$  is  due to the curvature of $AdS_5$. 
Since $S$ should take integer values  at  the quantum level, 
this suggests \gkp, by analogy with the large R-charge case \bmn, 
 that this string state  should correspond to a CFT operator with large canonical 
dimension equal to $S$.  
A remarkable  observation  made in \gkp\ is that 
the relation $E \approx S + ...$ contains also a subleading logarithmic  term, 
\eqn\subbe{ \E_{_{\S \gg 1}}  \approx  \S +   { 1 \ov \pi} \ln \S + ... , \ \
\ {\rm i.e.} \ \ \ \ \
 E_0  \approx S +  a_0 { \sql} \ln {S\ov \sql}  + ...   \ , \ \ \ \ \ \ \ \ 
a_0 =  { 1 \ov \pi} \ .  } 
This is an  example of  the general relation \cla. 
As another example, let us mention that 
 for the string oscillating in $AdS_5$ \drew\  
one finds \rf{\ves,\min}\ ($\cN \gg 1$):
$\E  \approx\cN + c_0 \sqrt{w \cN} + ... , $ i.e.
$E\approx \rN + c_0 \sqrt{ \sqrt \l  w \rN} + ...$.
For
 the string oscillating in $S^5$ \min:
$\E\approx\cN + c_1 {w^2 \ov  \cN} + ... , $ i.e.
$E\approx \rN + c_1  {\l w^2 \ov  \rN} + ... $.

Eq. \subbe\  may be compared \gkp\ to the known similar behaviour \old\ of the  
anomalous dimension  of the dual gauge theory operator, 
e.g. Tr$(\Phi^* D_{i_1} ...D_{i_S} \Phi)$, 
in an asymptotically free theory  (i.e. near a UV fixed point). 
Since the  rotating string state is not  BPS, the relation \subbe\ 
is expected to be  renormalized by  quantum corrections  on both string-theory 
 and gauge-theory sides.  On the gauge-theory side, it is believed (see \old\    and 
discussion in \gkp) 
that the anomalous dimension should  not contain higher powers  of $\ln S$ 
(even though they appear in individual graphs). 
Fortunately,  one is able to argue (see \frt\   and below) 
that  the same is true also  on the string-theory side:
there is a simple scaling argument that implies that  all sigma model 
$\a' \sim { 1 \ov \sql}$ corrections  to  the energy of the rotating string 
solution \subbe\ do not produce higher than  first powers  of $\ln S$ in $E$, 
i.e. they modify \subbe\  only by an ``interpolating  function'' of  the 
string tension (i.e. of  the 
`t Hooft coupling) 
 \eqn\subb{ E_{_{ {S\ov \sql} \gg 1}}
  \approx S +   f(\l)  \ln {S}  + ...   \ ,
\ \ \ \ \ \  \   \  
f(\l \gg 1) = a_0 {\sql}  + a_1 + { a_2 \ov \sql} + ...    \ .  } 
This is the same kind of 
 modification due to quantum corrections  that happens  on the gauge-theory  side, 
where  $f( \l \ll 1) = b_1 \l + b_2 \l^2 + ... $. 

The starting point of the semiclassical 
quantization is again the GS action in \adss  on a 2-d cylinder 
expanded  near the solution 
\spin. This is a conformal  2-d theory \MET, i.e. 
should be  no 2-d UV divergences
in expansion near any classical string solution.
One needs to impose \frt\ the  angular momentum quantization 
condition  $ \langle \Psi| \hat S |\Psi \rangle  = S= \sql \S$=integer  and compute 
$ \langle \Psi| \hat E |\Psi \rangle  = E(S)$.  The semiclassical 
string spectrum contains the ground state 
$|\Psi_0\rangle $ representing  the rotating string solution,  plus a tower of
 excited string  modes (corresponding to small oscillations on top of the 
macroscopic rotation). In contrast to the point-like $S^5$-rotating solution
discussed above, here the 
space-time (and effective 2-d) 
supersymmetry is broken, and thus there should be a non-zero quantum 
correction to the  ground-state energy, 
$ \langle \Psi_0| \hat E|\Psi_0 \rangle  - E_0 \not=0$.

The bosonic part of the quadratic  fluctuation  Lagrangian  can be
put into the form \frt\ 
$$L_{2B} 
= -  (\del \td t)^2   - \m^2_t \td t^2
+   (\del \td  \p)^2     + \m^2_\p \td \p^2
+ \ 4  \rr ( \k \sinh \r \ \del_0 \td t
-  w \cosh \r \ \del_0  \td \p ) 
$$  \eqn\onn{
+  \  (\del \rr)^2     + \m^2_\r  \rr ^2 
  +(\del \td \b_i)^2     + \m^2_{\b} \td \b_i^2
+ ( \del \td \psi_p)^2  \ ,  } 
where $\beta_i (i=1,2)$ and $ \phi\equiv \b_3 $ are the  $S^3$  angles of $AdS_5$
(see \add),  and $\psi_p$ are fluctuations of  5 angles of $S^5$.
Two out of 10 fluctuations may be eliminated using the 
conformal-gauge constraints. 
This action describes a mixed system of 2-d fields with non-constant ($\s$-dependent)
masses:
\eqn\maas{
\m^2_t  = m^2(\s)   - \k^2 
\  , \ \ \
\m^2_\p  =  m^2(\s)   - w^2 \ , \ \ \ 
 \m^2_\r  = m^2(\s)   - \k^2 - w^2 \ ,\ \ \ \ \ \m^2_\b  = m^2(\s) \ ,  }
$$ 
 m^2  \equiv 2 \r'^2 = 2\k^2 \cosh^2 \r - 2w^2 \sinh^2 \r \  . $$
 The quadratic fermionic action can be found from \fer,\form.
After applying  (as in \dgt) a particular $\s$-dependent fermionic rotation
$\vtt^I = U(\s) \theta^I$ and imposing the $\k$-symmetry gauge $\vt^1=\vt^2$ 
one gets an action describing 4+4 set of   2-d fermions  with a $\s$-dependent mass 
  as in \maas\ \frt:
\eqn\maan{
L_{2F} = i \bar \vtt \tau^a \del_a \vtt +  \bar \vtt  M  \vtt \ , 
\ \ \ \ M =  i \G_{234} m_F (\s)  \ , \ \ \ \ 
m_F = \r'(\s)  \  . }
``Diagonalizing'' the bosonic fluctuations,  it is possible to show the validity of 
the  sum rule\foot{Its existence reflects the fact 
 that the effective 
 2-d supersymmetry is broken spontaneously  by the classical solution.}
 $\sum m^2_B - \sum m^2_F=0$, 
which explicitly checks the UV finiteness of the 2-d theory.  
The quantum  correction to the ground-state energy is given (as in \asd)
by the expectation value of the 2-d Hamiltonian corresponding to \onn,\maan\
\eqn\eww{
E = E_0 + E_1 + ... = S    + {1 \ov \k} E_{\rm 2d} \   , 
\ \ \ \ \ \ \ \ \ \     E_{\rm 2d}=    {1 \ov \k} \langle 0| \hat H_{\rm 2d}|0\rangle \ . }
To find the   quantum correction  in the ``long-string'' limit 
one notes  that for $\S \gg 1$ 
 the masses of the 2-d fluctuation fields 
are approximately constant, $\r' \approx \k$,  and 
$\k(\S \gg 1) \approx  { 1 \ov \pi} \ln \S \gg 1$.\foot{Though the masses do change
 near the end-points,  one is able  to argue
  that since the change
 occurs only during a very short interval   of $\s$,
 this does not,
 for large $\S$,  significantly influence  the spectrum of the
 Laplace operators.}
Then  the 1-loop correction $E_1$ to the ground-state energy becomes \frt
\foot{For constant masses the expression for the quantum correction to
 the energy  becomes essentially the same as
  in section 6.5  of  \dgt.
  Here we consider the theory on a flat
 cylinder $(\tau,\s)$, so  that  the  individual field
 contributions to the 2-d vacuum energy are  the same
 as in the theory  with the  kinetic term $-\del^2 + m^2$.
 This is an analog  of the 2-d vacuum energy
$ (D-2) \sum_{n=1}^\infty n =  (D-2) \zeta(-1)
= - {D-2\ov 12} $ in the closed bosonic string case. It 
 vanishes in the standard flat-space   GS string case
 when all masses are  set to zero.}
 \eqn\vacc{ E_1  \approx   { 1 \ov 2 \k }
	\sum_{n=1}^\infty
\big[ \sqrt{n^2 + 4 \k^2  } + 2 \sqrt{n^2 + 2 \k^2}
+ 5 \sqrt{n^2} - 8 \sqrt{n^2 + \k^2} \ \big] 
 \ .    }
The large $\k$ (large mass) 
asymptotics is then 
  $E_1 =  { 1 \ov 2 \k } [ - { 3 \ov 2} \ln 2\  \k^2 + O(\k)]$, i.e.
\eqn\setw{
E_1 ( \S \gg 1)  \approx  a_1  \ln \S  \  , \ \ \ \ \ \ \  \ \ \ \ \ 
a_1 = -{3 \ln 2 \ov 4 \pi}  \ .   } 
This is consistent   with \subb, i.e.  there are no 1-loop quantum corrections 
that grow  faster than $\ln \S$ for large $\S$.

%%%%%%%%%%%%%%%%%%%%%%%%%%%%%%%%%%%%%%%%%%%%%%%%%%%%%%%%%%%%%%%%
Let us now argue \frt\ that  the same should be true also at
 higher orders in  inverse string  tension 
  expansion, i.e. that for large $\S$ \  ($\k \approx { 1 \ov \pi} \ln \S, \ \ \S \gg 1$)
\eqn\stoo{
E_l = { 1 \ov \k} ( E_{\rm 2d} )_l \approx 
 { 1 \ov \k  (\sqrt \l)^{l-1}} [ b_l \k^2 +  O(\k) ] \approx  
{1 \ov  (\sqrt \l)^{l-1}}  a_l \ln \S  + O(1)  \   . }
 The  2-d quantum field theory in question contains a collection
 of massive fields  with   approximately constant
 (for large $\S$) masses 
 proportional to $\k$ and  with interaction  terms containing
 two derivatives  or two powers of $\k$.
 It may  be modelled by a  Lagrangian (cf. \fla) 
 \eqn\modee{
L\sim \   \sqrt \l\  \big[ (1 + n_1 \xi^2 + n_2 \xi^4 +  ...)  (\del \xi)^2  +
 m^2 ( \xi^2 + k_1 \xi^4 + ...) \big]  \ , \ \ \ \ \ \ \ \  m\sim  \k \ , 
 }
 supplemented by massive 
 fermionic terms  which should lead to  cancellation 
of  all 2-d UV divergences.
  The theory is defined on the cylinder $0 < \tau < T, \ \ 
   0 \leq \sigma < 2\pi L $,   where  $T\to \infty$
   and $L$ is fixed (in the above discussion we put  $L=1$).
  Since this theory must be UV finite,  dimensional
  considerations  imply that
  the  $n$-loop contribution to the
  corresponding 2-d effective  action
 (and thus  to the  vacuum energy)  should  scale as
  \eqn\vaa{
\G_l= T ( E_{\rm 2d})_l  = {1 \ov  (\sqrt \l)^{l-1}}
  V_2  m^2  Q_l ( mL) \ , \ \ \ \ \ \ \   V_2 = T L  \ ,  } 
  where 
 $Q_l$ is a finite function of dimensionless ratio
  of the two IR scales $ 1/m$ and $L$.\foot{In  our  case of several fields
  with different masses  the functions  $Q_l$  will  also
  contain  finite ratios  of the  masses $\sim \k$, but these will
   stay constant  in the limit of large $\k$.}
Since in  the infinite volume limit $L\to \infty$ the
function  $Q_l$ should approach a finite constant $c_l$,
the same should be true also for fixed $L$ but for large $m$, i.e.
$Q_l(mL\gg 1) \to  c_l$. As a result,  for large $m \sim \k$
 we get (setting  $L=1$) 
 \eqn\asz{
 (E_{\rm 2d})_l \approx{1 \ov 
 (\sqrt \l)^{l-1}}   c_l    m^2  \approx  
{1 \ov  (\sqrt \l)^{l-1}}   b_l    \k^2 \  , }
thus  confirming \stoo. 
This implies that string corrections to the energy
of the rotating string solution
will  produce a  non-trivial  function $f(\l) $
in \subb\    but  will not lead to faster-growing $\ln^k S$ terms.  
Combined  with the  expected  absence
of the higher-order  $\ln^k S$ terms in the anomalous dimension 
of the minimal twist operators on the gauge theory side \rf{\old,\gkp}
this represents   a new non-trivial check of the gauge
theory -- string theory correspondence.

\bigskip 

%%%%%%%%%%%%%%%%%%%%%%%%%%%%%%%%%%%%%%%%%%%%%%%%%%%
\subsec{\bf String rotating in $AdS_5$ and boosted in $S^5$ }
%%%%%%%%%%%%%%%%%%%%%%%%%%%%%%%%%%%%%%%%%%%%%%%%%%%%%%%%%%%%

Let us now consider the classical solution  that
generalizes the above two, i.e.   describes 
a folded closed  string rotating  in $AdS_5$   with its center of mass 
moving along  the big circle of $S^5$  \frt\  (cf. \poi,\spin)
\eqn\poin{
G: \ \ \  t= \k \t \ , \ \ \ \p = w \t \ ,
\ \ \  \vp = \n \t  \ , 
 \ \ \r= \r(\s) \ , \ \ 
 \ \  \r'^2 = \k^2 \cosh^2 \r - w^2 \sinh^2 \r - \n^2   \ . }
 The corresponding string modes
will carry two global charges,\foot{The classical parameters are again related by the $\s$-periodicity condition \frt:  
$w=w(\S,\n)$, $\k=\k(\S,\n)$, etc.} 
 $J= \sql \nu$ and $S=\sql \S$
with  the energy  being function of them $E=E(J,S)$. They will be counterparts  
of  gauge-theory operators  which have 
large  spin and large R-charge at the same time.
The expression for the classical 
 energy  $E=E(J,S)= \sql \E (\nu, \S)$
will contain both $E=J$ and $E=S + a_0 \sql \ln S + ... $ cases 
as special limits.

One interesting aspect of this solution is that it illustrates 
the interpolation between expansions near different semiclassical points.
For example, 
in the limit of large $\n$ and 
small $\S$ the dependence of the classical energy  $E=E(J,S)$
on $S$ will be the same as in the quantum oscillator part of \rell\ 
corresponding to point-like $J\not=0$ solution: as in flat space, 
the classical spin can be built out of quantum-oscillator spins.
The agreement  between  the classical and quantum expressions here is non-trivial 
since it depends on  the curvature of $AdS_5$.  

In the  ``short-string'' limit one has 
$(2 \S)^2 \ll 1 + \n^2$, and 
 finds different expressions for $E$ depending on  values of $\n$.
If $\nu \ll 1$ then $\S \ll 1$ and  we get  
the same relation as in flat space: 
\eqn\jjj{ E\approx  \sqrt { J^2  + 2 \sql S }    \ . }
Indeed, 
 $J$ has the meaning of  a   linear  momentum along the $\vp$-direction 
of $S^5$  and $S$ is the spin
of a state on the leading Regge trajectory.  
If $\nu \gg 1$ (so that $\nu \gg 2 \S$) then 
\eqn\ssj{
E \approx  J + S + { \l \ov 2 J^2} S + ...    \  . }
This is to be compared with the quantum-corrected energy 
of the point-like solution  with large $J$ \rell,\relol.
For the oscillator states on the leading Regge trajectory
 $(a_1^\dagger a^\dagger_{-1})^{S\ov 2} |0\rangle$  one has
 $n=\pm 1, \ N_1=N_{-1}= {S\ov 2}$  and thus 
\relol\ takes the form  
$E \approx  J + (1  + { \l \ov 2 J^2} + ... )  S + ... $ 
which is in perfect agreement with \ssj. 
This provides a non-trivial check\foot{Notice again that the presence of the 
$S$-term  in both \ssj\ and  the expression following from 
 \relol\  is due to the curvature of $AdS_5$.}
 of consistency  of the semiclassical expansion. 
For $ J \gg S$ the  corresponding gauge-theory  operators 
 \rf{\bmn} should be \rus: 
$\sum_{_{l_1,...,l_S=1}}^J {1\over \sqrt{J} N^{J/2} }
\ {\rm Tr}\big(...ZD_{i_1}Z...ZD_{i_S}Z...\big) e^{ {2\pi i\over
J}  (l_1+...+l_S) }$, i.e. they   should 
contain $J$ factors of $Z$-scalar and $S$ covariant derivatives 
(at positions $l_k$)
so that  their  canonical dimension is  $\Delta_0= J + S$.
As was noted  in \rus, the $O({ \l \ov 2 J^2})$ correction to 
the anomalous dimension of  such operators should  indeed appear 
from a perturbative computation on the SYM side, 
in agreement with string-theory 
result \ssj.

In the  ``long-string'' limit  the spin parameters  is always large, $\S \gg 1$.
If $\nu \ll \ln \S$  we get the  expression  similar to the one in the $J=0$ case 
\subbe
\eqn\sedl{
E \approx  S + { \sql  \ov \pi} \ln { S \ov \sql} + 
 { \pi J^2 \ov 2 \sql \ln { S \ov \sql} }  + ... \ ,  }
while if $\n \gg \ln \S$ 
\eqn\sedo{
E \approx  S +  J  + {  \l \ov 2 \pi^2 J } \ln^2 { S \ov J} + ...\ ,  }
i.e. for large enough $J$ (or R-charge) we
 no longer find the $\ln S$ term \frt. 

As in the previous  special $S=0$ and $J=0$  cases, one can  again 
develop the semiclassical quantization  of this solution \frt, 
finding the quantum correction to 
the ground state energy (cf. \asd,\eww) as well 
the  energies of the tower of string oscillator states that have the same $S$ and $J$ 
quantum numbers but different 
oscillator occupation numbers. 

Other  similar ``non-topological soliton''-type 
solutions  for strings in \adss were discussed in \rf{\ves,\gkp,\rus,\min}.
Different  classical solutions ``probe''
different types of string motions, i.e. different sectors of the 
complete 
string spectrum,  and their further study should be useful.

%%%%%%%%%%%%%%%%%%%%%%%%%%%%%%%%%%%%%%%%%%%%%%%%%%%%%%%%%%%%%%%%%%%%%%%%
%%%%%%%%%%%%%%%%%%%%%%%%%%%%%%%%%%%%%%%%%%%%%%%%%%%%%%%%%%%%%%%%%%%%%%%%

%%%%%%%%%%%%%%%%%%%%%%%%%%%%%%%%%%%%%%%%%%%%%%%%%%%
\newsec{Near-conformal (near-AdS) cases }
%%%%%%%%%%%%%%%%%%%%%%%%%%%%%%%%%%%%%%%%%%%%%%%%%%%%%%%%%%%%

It is of obvious interest to try to generalize the semiclassical 
approach to string -- gauge theory duality to non-conformal 
(and less supersymmetric) cases. For example, the $S + f(\l) \ln S$ 
form of the dimension of minimal twist operators  
should be  universal, i.e. it should be found  \gkp\  in   all 
(e.g.,  non-supersymmetric, asymptotically free) 
gauge theories  near a (UV)  fixed point.  
It is then natural to try to reproduce this  behavior on the 
string theory  side  for   non-conformal examples of duality
like the one in \rf{\KS}. 
The notion  of (anomalous) dimension
is defined only near a conformal fixed point, 
so the  semiclassical approach as described in \rf{\gkp,\frt}
and above  should directly 
apply only in the  ``near-AdS''
region. The  UV   region of the corresponding 
 background    is described  by the type IIB supergravity 
 solution of 
 \KT\ 
representing fractional D3-branes 
 on a conifold 
($N$ D3's  plus  $M$  D5's  wrapped on a 2-cycle):
its UV (large distance)  asymptotics   is  approximately   $AdS_5 \times \ton$.

In trying to repeat the  analysis 
done above for the \adss case in the  near-conformal case one faces two 
problems. The first is to   find  the form of the corresponding  classical string solutions. The non-conformal  backgrounds are naturally 
written  in the Poincare-type coordinates (more precisely, their 
$AdS$  limit gives the $AdS$ metric in Poincare coordinates), 
while the  rotating string solutions 
in \adss  are  simplest in the global coordinates
(cf. \spin,\pin).
 Here 
 it will be  useful to be guided by  
 the form  of the \adss  solutions in the Poincare coordinates 
\poo,\pin. 

The second (related) problem is  how to define the string-theory analog 
of conformal dimension, given that  the background is only asymptotically $AdS$. 
In particular, the  global-coordinate energy $E$  will no longer be conserved.
On the gauge theory side, this 
 may be viewed as a reflection of running of gauge  couplings
(which enter the expression for anomalous dimension) once one moves 
 away from the  conformal point.\foot{
It may be possible to relate $\del E \ov \del \tau$ 
to an analog of RG equation on the gauge theory side.}  
Below we shall  use the same  definition of $E$ as in the 
$AdS$ case 
expressed in terms of the Poincare-coordinate  fields \eee.
The definition of  dimension or $E$ away from the conformal point is of
course ambiguous, but viewing the  geometry of \KT\ as a perturbation 
of  $AdS_5 \times \ton$, 
this ambiguity  should not matter  to  leading order in 
deviation from the conformal point.
%\foot{Here we  follow the  standard
%logic of conformal perturbation theory.}

Our aim will be to find the correction to the string energy 
$E$ (and thus to  the corresponding dimension) 
due to non-conformality  of the background. 
We shall consider both the case of the point-like string 
moving along a circle in $\ton$
and a folded string rotating parallel to the boundary. 

%We shall find that  in the first case 
%the leading ``non-conformal''  correction to $E$ 
% vanishes; this may be related to the similar result in 
%\fkt, and  should  have a simple gauge-theory explanation.

%But main aim is to compute  correction to the rotating solution 
%to leading order in parameter of deformation of $AdS_5$ in the KT direction.

\bigskip

%%%%%%%%%%%%%%%%%%%%%%%%%%%%%%%%%%%%%%
\subsec{\bf Perturbed  solution for point-like string  rotating in $\ton$}
%%%%%%%%%%%%%%%%%%%%%%%%%%%%%%%%%%%%%

Our starting point will be  the following generalization of the 
$AdS_5 \times S^5$ metric
\eqn\asa{
ds^2=  h^{-1/2}(y)  dx_m dx_m +   h^{1/2}(y) ( dy^2 + y^2 ds^2_{5})     
 \ , } 
 \eqn\seda{
 h= {Q(y) \ov y^4} \ ,\ \ \ \ \  \ \ \ \  Q =   1 + \ee  q(y) 
\ . } 
Here $\ee=0$ corresponds to the  conformal case, 
e.g.,  $AdS_5 \times \ton$   if $ds^2_5= ds^2_{\ton}$.
An  example of \asa\ 
 is the solution of \KT\  where (after appropriate rescaling)\foot{This 
metric 
is supported by certain   p-form fluxes, which will not be important 
at the classical level 
for the  solutions we discuss below.}
\eqn\wwq{    q = \ln  y  \ , \ \ \ \ \ \ 
\ \ \    \ee  = { 3g_s M^2 \ov 2 \pi  N }  \ . }
Assuming  $\ee \ll 1 $  and the range of $y$ 
  such  that $\ee \ln y$ is small,   this  metric is close to 
$AdS_5 \times {\ton}$ and can be written as (expanding in powers of $\ee$) \foot{In terms of global coordinates of $AdS_5$ 
this metric can be written as 
  $$ds^2 = Q^{-1/2} (y)
 [ - {\rm cosh}^2 \r \ dt^2 +  d\r^2 + {\rm sinh}^2\r \ d\Om_3 ]
+ Q^{1/2} (y) ds^2_{\ton} +  {dy^2 \ov y^2} [ Q^{1/2} (y) - Q^{-1/2} (y) ]\ , $$  
where $y=y(\r,t,\beta_i)$. It is clear that for $Q \not=1$ 
the correction to the metric starts depending on $t$.}
$$ds^2=  {1 \ov z^2} [ 1 - \ha \ee q(z) ]  dx_m dx_m +   
[  1 +  \ha \ee q(z) ]  {dz^2 \ov z^2}       +    [  1 +  \ha \ee q(z) ]  ds^2_{\ton}    + O(\ee^2)   $$
\eqn\asar{
= \  {1 \ov z^2} (  dx_m dx_m +   {dz^2 })   +  ds^2_{\ton} 
+    \ha \ee q(z) \big[  {1 \ov z^2}(  - dx_m dx_m +   {dz^2})   +  ds^2_{\ton} 
\big]
 + O(\ee^2) \  , } 
\eqn\seat{    z\equiv  { 1 \ov y}  \ ,\ \ \ \ \ \ \ \ \ q(z)  = - \ln z  \ .    } 
The  string action on this background 
may thus be   interpreted as  perturbed string action  on 
 $AdS_5 \times \ton$.

Let us now find  the generalization of the  solution \poo\  to the leading order 
in $\ee$ (now $\vp$  is an angle of ${\ton}$).\foot{In the  ``plane-wave''  
context this  null geodesic in    $AdS_5 \times \ton$  
 was considered in \rf{\kll,\oog,\ppz}.}
Assuming the ansatz \ $x_0 = x_0 (\t),  \ z=z(\t), \ \vp= \vp(\t)$  with all other coordinates being  zero or constant   the perturbed solution corresponding to \asar\ 
should have the form (we set $\n=1$)
\foot{The equations for $x_0$
and $\vp$  are directly integrated while the equation for $z$ is given by the constraint $ - {1 \ov z^2} [ 1 - \ha \ee q(z) ]  \dot x_0^2    +  
[  1 +  \ha \ee q(z) ]  {\dot z^2 \ov z^2}       +    [  1 +  \ha \ee q(z) ] \dot \vp^2  =0 $, i.e. 
$       
[  1 +  \ha \ee q(z) ] ( - z^2 +  {\dot z^2 \ov z^2} ) 
      +    [  1 -  \ha \ee q(z) ] 
 + O(\ee^2)  =0 $.  The explicit solution for  $Z$ is
($q= \ln \cos \t$): 
$Z(\t) = - \ha \tan \t \ \int^\t_0 d\t' \cot \t' \ln \cos \t'  $, 
i.e.  it can be expressed in terms of a polylog function. 
  For  general  $Q(y)$   the equations   for the corresponding null geodesic are 
$y^2 \dot x_0 = \k Q^{1/2} (y) \ ,\ \  \phi = w Q^{-1/2}(y) \ , \ \ 
\dot y^2 + w^2 y^2 Q^{-1} (y) - \k^2 =0 $  (see also \ppz). }
\eqn\sea{
x_0 = \tan \t  + \ee T (\t) \ ,\ \ \ \  
z = {1 \ov \cos \t} [1 - \ee  Z(\t)] \ ,\ \ \ 
\vp = \t + \ee \Phi (\t)  \ ,   }
\eqn\duy{
\cos^2 \t \ \dot T  = -2 Z + \ha F \ , \ \ \ \ \
\dot \Phi = - \ha F \ ,   \ \ \  \  \ 
\tan  \t \dot Z =  { 1 \ov \cos^2 \t}  Z  - \ha F  \ ,  \ \ \ \ \ \
 F(\t)\equiv q(z(\tau)) \ . } 
The above relations are valid for arbitrary $q(z)$. 
Computing  the value of the  energy density  (defined as 
in the $AdS$ case by  \sre)  
 on the deformed solution of the form \sea \ we get 
\eqn\eqe{
\E_d = 1 +  { \ee \ov \cos^2 \t} (  Z + \sin \t  \cos \t \ \dot Z + \cos^2 \t \dot T )  + O(\ee^2) 
\ . }
It then  follows from the form of the solution in \duy\   that 
the leading $O(\ee)$ term in $E$ cancels out, 
$
\E_d = 1   + O(\ee^2) $, i.e. 
\eqn\serr{
E=J + \ee \g_1 (\tau) + \ee^2  \g_2(\tau) + ...
 \ , \ \ \ \ \ \ \ \  \g_1 =0  \ . }
Note that  $\g_1=0$ for any  $ F(\tau)$, but 
  higher-order corrections certainly should  not  vanish.

Thus 
 there is no leading  $O ({M^2 \ov N})$ correction to the  anomalous 
dimension of the corresponding ``ground-state''  large R-charge 
operator  Tr$(A_2 B_2)^J$  \kll\  due to running 
of the gauge couplings in  the  dual $SU(N) \times SU(N+M)$ 
gauge theory. 
This  is reminiscent of 
the  vanishing of the leading correction to the anomalous  dimension 
of the operator 
Tr$(A_i B_j)$ observed in \fkt\ 
 -- the correction to the dimension starts at a rather high order in $M/N$: 
$\Delta = -  \ha   + c_1 { \zeta(3) \ov \sql} ({M \ov N})^4 + ...$.
It would be interesting to find a  field-theory 
explanation of 
  why the dimension
of the chiral operator  Tr$(A_2 B_2)^J$ of 
the conformal $SU(N) \times SU(N)$ theory   \KW\  
 does not shift  if we make
an $O(M^2)$   marginal perturbation  away  
from the conformal point.\foot{ The same may   be true 
also for the  LS \LS\  perturbations of $\cal N$=4 SYM theory.
The supergravity dual of large R-charge limit of LS conformal point  
was considered in \rf{\lei,\bre}.
%  mostly  the LS conformal point was considered, not 
%full RG flow leading to it. 
Let us note also that a  discussion of a plane wave  limit of a  solution 
 describing flow between  two conformal points 
was given in   \rf{\gimo,\bre}. }
%In general, dimension  is  $ < O(x) O(x') > \propto  (|x-x'| \mu)^\Delta $, 
%$\Delta$ is const, but now it will also run with $\ln \mu$.

\bigskip

%%%%%%%%%%%%%%%%%%%%%%%%%%%%%%%%%%%%%%
\subsec{\bf Perturbed  solution for rotating  string}
%%%%%%%%%%%%%%%%%%%%%%%%%%%%%%%%%%%%%
 
Let us now  carry out a similar computation for  the spinning string solution
\pin. 
The aim  is to see how  the   $E(S)$ relation \subbe\ found in the $AdS_5$ case 
is modified    when the $AdS_5$ the metric  is replaced by 
its  deformation in \asar.
 The internal $\ton$  part of the metric  will 
not be important as the corresponding angles will be kept fixed. 
Here we expect a non-trivial correction to $E$ due 
to the deformation, 
\eqn\drew{
E= S + [ a_0 {\sql}  + \ee \eta_1  (\tau) + O( \ee^2) ] 
\ln S + ...\
, }
where $\tau$-dependence of the function 
$\eta_1$  should be reflecting broken  conformal invariance 
 (i.e. changing of anomalous dimension with scale 
due to running of gauge couplings).\foot{The dependence on $\tau$ may be traded for the dependence on $z$ that plays the role of an energy scale.}

It is useful to change  the coordinate $\s$  to  
$ s = \tanh \r$, where $\r(\s)$ is a solution of the condition in \spin\ 
(we set $\k=1$):
\eqn\cha{
({ds \ov d \s})^2 = ( 1 - s^2) (1 - w^2 s^2) \equiv G^2 (s)  \ .}
The relevant part of the string action  in conformal gauge is then 
$$ I \sim \int d \t ds  \bigg[  z^{-2} (1 + \ha  \ep \ln z)   G^{-1} (
 - \dot x_0^2  + \dot r^2  + r^2 \dot \p^2 ) 
$$
\eqn\laa{
  -    G  ( 
 -  x'^2_0  +  r'^2  + r^2  \p'^2)
+  z^{-2}  (1 - \ha  \ep \ln z) 
 (G^{-1}  \dot z^2     -    G  z'^2 )  \bigg] \ , 
}
where prime  now stands for the derivative over $s$.
The  $AdS$  rotating solution  \pin\ is then generalized, to leading order in $\ee$, 
as follows
$$ 
 x_0 = \tan \tau  + \ep  T(\t,s)  \ , \ \ \ 
 z= { \sqrt{ 1- s^2 } \ov \cos \tau } [ 1  - \ep   Z(\t,s) ]\ , 
$$  \eqn\pink{  
 \p = w \t  + \ep  P(\t, s)  \ , \ \ 
 \  \  r= { s \ov  \cos \tau } [1 + \ep  R(\t,s) ]  \ .  }
The aim is  to find the perturbed solution, 
and then  to compute the leading correction to the 
energy density \sre\ (where $x^2= - x_0^2 + r^2$). 
 In general, we get (cf. \eqe)
$$\E_d= {1 \ov 1-s^2} \bigg(  1 +  
{ \ep \ov \cos^2 \t}
\big[(1 +     s^2 \cos 2 \t )   Z   +   s^2 \cos 2 \t  R 
$$   \eqn\eqel{
+ \cos^2  \t\ \dot T -  s^2 \sin \t \cos \t   \dot R +      (1- s^2)
 \sin \t \cos \t    \dot Z \big] \bigg)  + O(\ep^2) 
\ .  }
Note that  in \pin, i.e. at the zeroth order in $\ep$,  
  $ \del_\t ( { z \ov r} ) =0$. Assuming that this will be  true also at non-zero  
order, i.e. that  $\dot Z + \dot R =0$, one is able to separate $\tau$ and $s$ 
coordinates in the solution, 
$
T= T(\t) ,   \ \  P=P(\t)  ,  \ \
Z= A(\t) + B(s) ,\ \ R= - A(\t) + C(s)   . 
$
One is then to solve the resulting string equations,  
$
w( \tan \t + 8  \dot A ) + 2 \ddot P =0, \ 
  \tan \t  + 4 \dot A  
- 4 \sin \t \cos \t \dot T + 2 \cos^2 \t \ddot T  =0$, etc., 
and the conformal-gauge  constraints. 
%relating B and C 
Finally, one is to   compute the energy \eee,\eqel\ and the spin 
(the conserved charge  corresponding to translations in $\phi$ in \laa) 
and to compute $\eta_1(\tau)$ in  \drew. 
We leave detailed discussion of this computation for  future.

%In fact, we may assume that $x_0$ and $\p$ depend only on $\t$ so that then 
%$$\dot x_0 = z^2 (1 - \ha  \ep \ln z) , \ \ \ \ 
%\dot \p = w  z^2/r^2   (1 - \ha  \ep \ln z)$$

%%%%%%%%%%%%%%%%%%%%%%%%%%%%%%%%%%%%%%%%%%%%%%
\bigskip
\noindent
{\bf Acknowledgements}
%%%%%%%%%%%%%%%%%%%%%%%%%%%%%%%

\noindent

I am grateful to  S. Frolov for  a 
collaboration on \frt\ and  many discussions of other related 
issues mentioned above. 
 % other  results  presented   above.
I would like also  to thank 
 I. Klebanov,  R. Metsaev,  A. Polyakov, J. Russo  
and K. Zarembo  for  useful   discussions and comments. 
This work  was supported by the grants DOE 
DE-FG02-91ER40690,   PPARC SPG 00613,
 INTAS  99-1590,   
%EC grant HPRN-CT-2000-00131
 and by  the Royal Society  Wolfson Research Merit Award.

%%%%%%%%%%%%%%%%%%%%%%%%%%%%%%%%%%%%%%%%%%%%%%%%%%%%%%%%
\vfill\eject
\listrefs
\end

 higher orders of the  inverse string  tension 
  expansion, i.e. for large $\S$ \  ($\k \approx { 1 \ov \pi} \ln \S, \ \ \S \gg 1$)
\be\la{stoo}{
E_l = { 1 \ov \k} ( E_{\rm 2d} )_l \approx 
 { 1 \ov \k} [ b_l \k^2 +  O(\k) ] \approx  a_l \ln \S  + O(1)  \   . }\ee
 The  2-d quantum field theory in question contains a collection
 of massive fields  with   approximately constant
 (for large $\ss$, i.e. in the ``long string' case) masses 
 proportional to $\k$ and  with interaction  terms containing
 two derivatives  or two powers of $\k$, 
  \be{
L\sim  \sqrt \l \big[ (1 + n_1 \xi^2 + n_2 \xi^4 +  ...)  (\del \xi)^2  +
 m^2 ( \xi^2 + k_1 \xi^4 + ...) \big]  \ , \ \ \ \ \ \ \ \  m\sim  \k \ ,  
 }\ee
 supplemented by massive 
 fermionic terms  which should lead to  cancellation 
of  all 2-d UV divergences.
  The theory is defined on the cylinder $0 < \tau < T, \
   0 \leq \sigma < 2\pi L $,   where  $T\to \infty$
   and $L$ is fixed (in the above discussion we had $L=1$).
  Since this theory must be UV finite,  dimensional
  considerations  imply that
  the  $n$-loop contribution to the
  corresponding 2-d effective  action
 (and thus   vacuum energy)  should  scale as
  $
\G_l= T ( E_{\rm 2d})_l  = {1 \ov  (\sqrt \l)^{l-1}}
  V_2  m^2  Q_l ( mL) $ where $   V_2 = T L $ 
  and 
 $Q_l$ is a finite function of dimensionless ratio
  of the two IR scales $ 1/M$ and $L$.
Since in  the infinite volume limit $L\to \infty$ the
function  $Q_l$ should approach a finite constant $c_l$,
the same should be true also for fixed $L$ but for large $m$, i.e.
$Q_l(mL)_{m \gg 1} \to  c_l$. As a result,  for large $m \sim \k$
 we get (after setting $L=1$)
 \be{
 (E_{\rm 2d})_l \approx{1 \ov 
 (\sqrt \l)^{l-1}}   c_l    m^2  \approx  
{1 \ov  (\sqrt \l)^{l-1}}   b_l    \k^2 \  , }\ee
 confirming \rf{stoo}. 
This implies that string corrections to the energy
of the rotating string solution
will  produce a  non-trivial  function $f(\l) $
in \rf{subb}    but  will not lead to faster-growing $\ln^k S$ terms.  
Combined  with the  expected  absence
of the higher $\ln^k S$ terms on the gauge theory side \ci{old,gkp}
this provides  a new non-trivial check of the gauge
theory -- string theory correspondence.

%%%%%%%%%%%%%%%%%%%%%%%%%%%%%%%%%%%%%%%%%%%%%%%%%%%
\newsec{Conclusions }
%%%%%%%%%%%%%%%%%%%%%%%%%%%%%%%%%%%%%%%%%%%%%%%%%%%%%%%%%%%%
New directions to compare gauge theory with string theory.
Need better understanding of string theory to derive non-trivial interpolating functions. 

GS action is a useful practical  tool, produces 
well-defined semiclassical expansion. 
Progress in  clarification of part of string spectrum in $(J,S)$ sector 
in and beyond leading order in $1 \ov \sql$.

Extension to non-conformal   and non-susy cases 
-- expected  universality of $E= S + \ln S$. 

Importance of \lc gauge.

Our aim is to see also how the correction depends on spin $S$ 
and eventually give a gauge theory interpretation to it.
In particular, how derivative $\dot E$ depends on  $S$.

Rotating solutions in Warner-Pilch   case ?
In the  L-S fixed point  all is trivial 
if we fix the angle  in the warp factor -- 
the same solution.